\documentclass[final,3p,times,twocolumn,10pt]{elsarticle}

\usepackage{amssymb}
\usepackage{amsmath}
 \usepackage{url}
\usepackage{forest}
\usepackage{dirtree}
\usepackage[version=4]{mhchem}

\usepackage{courier} 

\usepackage{listings}
\usepackage{xcolor}

\definecolor{codegreen}{rgb}{0,0.6,0}
\definecolor{codegray}{rgb}{0.5,0.5,0.5}
\definecolor{codepurple}{rgb}{0.68,0,0.82}
\definecolor{backcolour}{rgb}{0.985,0.98,0.972}
\definecolor{codeorange}{rgb}{0.90,0.5,0.0}
\definecolor{codeblue}{rgb}{0,0,0.5}

\lstdefinestyle{mystyle}{
    backgroundcolor=\color{backcolour},   
    commentstyle=\color{codegreen},
    keywordstyle=\color{codepurple},
    numberstyle=\tiny\color{codegray},
    stringstyle=\color{codeblue},
    basicstyle=\ttfamily\footnotesize,
    breakatwhitespace=false,         
    breaklines=true,                 
    captionpos=b,                    
    keepspaces=true,                 
    numbers=left,                    
    numbersep=5pt,                  
    showspaces=false,                
    showstringspaces=false,
    showtabs=false,                  
    tabsize=2
}

\journal{ArXiv}

\begin{document}

\begin{frontmatter}



\title{\textbf{aPriori: a Python package to process direct numerical simulations}}


\author[a,b,c]{Lorenzo Piu}

\author[c]{Heinz Pitsch}

\author[a,b,d]{Alessandro Parente}

\affiliation[a]{organization={Aero-Thermo-Mechanics Laboratory, Université Libre de Bruxelles, École Polytechnique de Bruxelles},
            addressline={Avenue Franklin Roosevelt 50}, 
            city={Brussels},
            postcode={1050}, 
            country={Belgium}}

\affiliation[b]{organization={Brussels Institute for Thermal Energy (BRITE)},
            city={Brussels},
            postcode={1050}, 
            country={Belgium}}

\affiliation[c]{organization={Institut für Technische Verbrennung, RWTH Aachen University},
            addressline={Templergraben 64}, 
            city={Aachen},
            postcode={52062}, 
            country={Germany}}

\affiliation[d]{organization={WEL Research Institute},
            addressline={Avenue Pasteur 6}, 
            city={Wavre},
            postcode={1300}, 
            country={Belgium}}

\begin{abstract}

In the field of computational fluid dynamics, direct numerical simulations generate highly detailed data for the analysis of turbulent flows by resolving all relevant physical scales. Yet their large size, complexity, and heterogeneity make systematic post-processing and data reuse increasingly challenging. Despite the growing availability of high-fidelity simulations through public repositories, extracting meaningful physical insight often requires substantial technical effort, specialized workflows, and access to high-performance computing resources. In this article we introduce \texttt{aPriori}, an open-source Python package developed to address these limitations by providing a dedicated, memory-efficient, and user-oriented framework for the analysis of direct numerical simulation data. The software enables streamlined handling of three-dimensional fields, including filtering, scale separation, gradient evaluation, thermochemical analysis, and visualization, using concise and reproducible scripts. Its pointer-based data management strategy allows very large datasets to be processed on standard workstations without excessive memory usage, significantly lowering the barrier to advanced analysis. Beyond basic post-processing, \texttt{aPriori} supports workflows central to modern turbulence and combustion research, such as \textit{a priori} model assessment, data-driven closure development, and detailed chemical analyses that include computational singular perturbation. By unifying these capabilities within a coherent and extensible software architecture, \texttt{aPriori} enhances productivity, promotes reproducibility, and facilitates broader and more effective use of high-fidelity simulation data within the computational fluid dynamics community.
\end{abstract}




\end{frontmatter}



\section{Motivation}\label{sec:Motivation}


Direct Numerical Simulation (DNS) has become a widely used tool for researchers studying turbulence and combustion due to its ability to solve the full Navier-Stokes equations without relying on subgrid-scale models \cite{moin1998direct}. DNS offers a direct resolution of all scales of turbulent fluid motion, from the largest energy-containing eddies down to the smallest scales where kinetic energy is dissipated into heat and, in reactive flows, reactants are mixed at a molecular level and find favourable conditions to react. These details provide insight into complex physical phenomena, such as the statistical properties of turbulence \cite{Pope2012-jn} and the interaction between turbulent mixing and chemical reactions in combustion processes \cite{PoinsotVeynante_2005}.
 
\begin{figure}[ht!]
    \centering
    \includegraphics[width=1.0\linewidth,trim=3 3 3 3,clip]{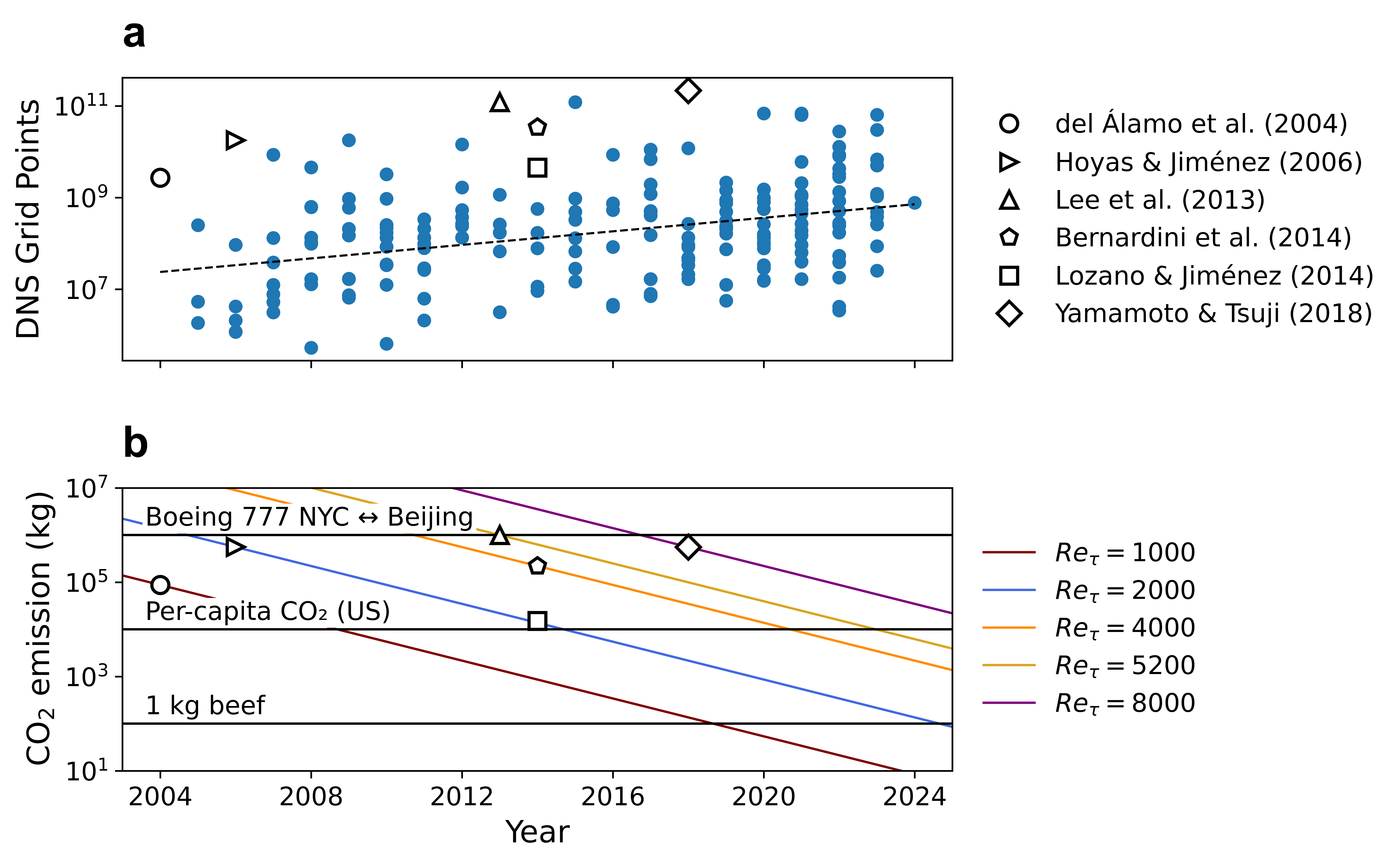}
    \caption{Estimated size (a) CO$_2$ emissions (b) associated with Direct Numerical Simulation (DNS) studies published in \textit{Journal of Fluid Mechanics} between 2004 and 2024. Each point represents a simulation reported in the literature, highlighting the rapid growth in computational cost---and therefore carbon footprint---of state-of-the-art DNS. The largest simulations exceed $10^3$ metric tons of CO$_2$, comparable to the emissions of a transcontinental commercial flight. 
    Coloured lines show model predictions for channel-flow DNS at fixed friction Reynolds numbers $Re_\tau$, based on a $Re_\tau^4$ scaling of computational cost and a Moore’s-law-like improvement in hardware efficiency over time.
    Source: adapted from \cite{Yang2024}.}
    \label{fig:DNS_CO2}
\end{figure}

The high level of detail in DNS reflects the substantial computational resources involved, which in turn results in significant \ce{CO_2} emissions. Yang \textit{et al.} \cite{Yang2024} analysed the DNS published in the Journal of Fluid Mechanics between 2004 and 2024, and estimated that the largest simulations can emit up to one thousand tons of \ce{CO_2}---the equivalent of a civil plane flying from New York to Beijing. Details are reported in Figure \ref{fig:DNS_CO2}. These authors also estimated that community efforts to share DNS data, such as the Johns Hopkins Turbulence Database (JHTDB) \cite{Li2008johnhopkins}, avoided roughly 3.3 million metric tons of carbon emissions, and this estimate underlines the importance of shared repositories that eliminate redundant computations.

Among these repositories, BLASTNet \cite{chung_wai_tong_2023_8034232, chung2022the, chung2023turbulence, CHUNG2022100087} stands out as a publicly accessible database recently released to support research in turbulent combustion. This archive, totaling 2.2 Tera Bytes, comprises 744 samples from 34 distinct high-fidelity DNS cases provided by multiple institutions. These cases encompass diverse flow types, including isotropic turbulence \cite{Chung2022, Donzis2010}, channel flows \cite{Samuel2024}, and different flame configurations \cite{Brouzet_Talei_Brear_Cuenot_2021, Coulon2023, Jung2021}. 
The availability of heterogeneous configurations is instrumental, among other applications, to train better machine learning algorithms, which are being increasingly studied to develop sub-filter closures \cite{swaminathan_parente_2023, Ihme2022-ac, Bode2019-ue, Bode2021-ur, Nista2023-re, deFrahan2019, Piu2025}.
Unlike benchmark datasets from the machine learning community (e.g. MNIST \cite{LiDeng2012mnist}, CIFAR \cite{Krizhevsky2009cifar}, and ImageNet \cite{Deng2009imagenet}), which provide ready-to-use configurations, extracting sub-filter quantities from DNS for modeling or retrieving chemical properties can require additional effort, eventually requiring notable computational resources for post-processing.

To tackle this, we propose \texttt{aPriori}, a Python package intended to simplify the workflows commonly involved in DNS data processing. The package adopts the data formatting convention used in the BLASTNet repository, which enforces a standardized structure that facilitates data sharing and reproducibility. This choice also enables the pointer-based data access strategy implemented in the library, while remaining sufficiently general to accommodate datasets that originate from different solvers provided they are converted to the required format.
\texttt{aPriori} includes built-in support for operations that are frequently required in turbulence and combustion studies, such as filtering scalar and vector fields for turbulent scales separation, computing gradients, and evaluating dissipation of scalar quantities. The integration of plotting utilities allows users to readily explore the results of the aforementioned operations, enabling faster interpretation and analysis. 

A key advantage of this package is its efficient memory management: instead of loading files into variables, it uses pointers to file locations, reading each file containing scalar variables only when accessed. This prevents memory overloads and facilitates the usability of large datasets. Moreover, it can be coupled with existing advanced chemistry packages to process reactive flows (e.g., Cantera \cite{cantera} and PyCSP \cite{Malpica2022pycsp}), and with Python-based libraries for machine learning and visualization (e.g., Pytorch~\cite{Paszke2019pytorch}, Tensorflow~\cite{tensorflow2015-whitepaper}, Matplotlib~\cite{Hunter2007matplotlib}, and Pyvista~\cite{Sullivan2019pyvista}). The software is accompanied by a documentation\footnote{\url{https://apriori.readthedocs.io}} and a dedicated GitHub repository\footnote{\url{https://github.com/LorenzoPiu/aPriori/tree/main}}, with access to the source code and to dedicated tutorials.

The manuscript is structured as follows. Section \ref{sec:software_description} presents the design principles and overall architecture of the \texttt{aPriori} package, describing its modular organization, the role and interaction of the core classes (\texttt{Scalar}, \texttt{Mesh}, and \texttt{Field}), and the adopted data formatting conventions required to interface with large DNS datasets. The section further details the implementation of key functionalities, including pointer-based data access, filtering and downsampling operations, gradient evaluation, data chunking strategies for thermochemical computations, and the integration with external chemistry and visualization libraries.
Section \ref{sec:usage_examples} illustrates the practical use of the software through a sequence of representative case studies and reproducible code snippets. These examples quantify the performance and memory efficiency of the pointer-based approach, demonstrate data loading and visualization workflows, and showcase typical \textit{a priori} analysis procedures for turbulence and combustion, including sub-filter quantities evaluation, model assessment, machine-learning–assisted closures, and Computational Singular Perturbation (CSP) analysis.
Finally, section \ref{sec:conclusions} summarizes the main contributions of the work, discusses the scope and limitations of the current implementation, and outlines prospective developments aimed at extending the library’s numerical capabilities, data abstractions, and applicability to a broader range of DNS post-processing tasks.

While the library can be installed and operate directly in high-performance computing (HPC) clusters, all tests presented in this work---which include processing of datasets with up to 1 billion grid points---were conducted on a local Unix-based system with an ARM64 architecture (8-core CPU, 16 GB RAM); the goal is to demonstrate the software’s capabilities to extend accessibility of DNS data beyond HPC environments.

\section{Software architecture}\label{sec:software_description}
The software architecture is modular, built around Python classes that represent key concepts like scalars, meshes, and fields in three-dimensional space. These classes allow users to compute various statistics, manipulate DNS fields, and visualize complex simulation data. The software’s modular architecture adheres to Python’s official packaging guidelines, enabling straightforward distribution through the Python Package Index (PyPI) \cite{pypi}. The following directory tree illustrates the main folders and the files structure of the software package:
\\
\dirtree{%
.1 aPriori.
.2 data/.
.2 dist/.
.2 docs/.
.2 LICENSE.
.2 pyproject.toml.
.2 README.md.
.2 src/.
.3 aPriori/.
.4 \_\_init\_\_.py.
.4 \_data\_struct.py.
.4 \_styles.py.
.4 \_utils.py.
.4 \_variables.py.
.4 derivatives.py.
.4 DNS.py.
.4 NN.py.
.4 plot\_utilities.py.
.2 tests/.
.2 tutorials/.
}

\begin{figure*}[ht!]
    \centering
\includegraphics[width=0.7\linewidth,trim=5 5 5 5,clip]{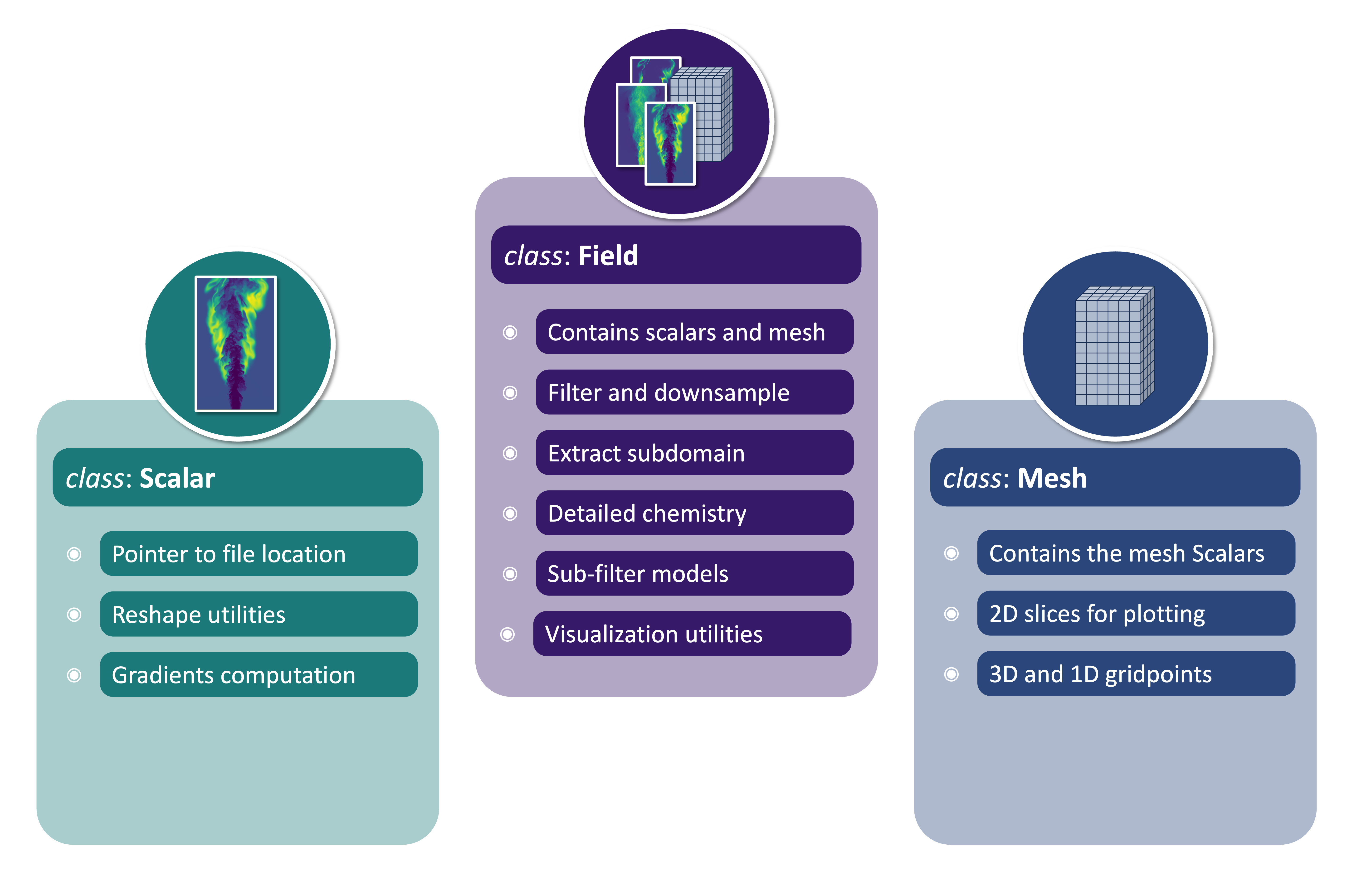}
    \caption{Graphical overview of the software architecture and the relationships between its three core classes. 
    The \texttt{Mesh} class stores the spatial coordinates and grid topology of the DNS domain, providing geometric information required for visualization and differential operations. 
    Individual physical variables (e.g.,\ velocity components, species mass fractions, pressure, temperature) are represented as \texttt{Scalar} objects, each of which handles memory-efficient access to the corresponding binary data files through a pointer-based mechanism. 
    The \texttt{Field} class acts as a high-level container that aggregates all \texttt{Scalar} objects associated with the same mesh, dynamically loading the variables present in the dataset, which are added as attributes of the class. 
    It provides unified methods for filtering, downsampling, gradient computation, thermochemical evaluations, and visualization.}
    \label{fig:architecture}
\end{figure*}

\vspace{8pt}
\noindent
The \texttt{data} directory houses example datasets employed to run the tutorials, while \texttt{dist} contains the files used for package distribution. The main source code is found under the \texttt{src/aPrioriDNS} directory, which is organized into modules. The \texttt{\_\_init\_\_.py} file contains code to initialize package-level variables and to import frequently used functions and classes. The \texttt{\_data\_struct.py} module defines the names of the main folders in which the dataset is required to be structured. The \texttt{\_styles.py} module controls the layout used in the \texttt{plot\_utilities.py} module. \texttt{variables.py} is the database containing all the available variables that can be part of the DNS field. The module \texttt{DNS.py} encompasses the three key classes \texttt{Scalar}, \texttt{Mesh}, and \texttt{Field}, which will be detailed later in this section. The \texttt{derivatives.py} module contains the functions for gradients and Laplacians computations, which are based on the \texttt{findiff} library~\cite{findiff}, and the \texttt{NN.py} module provides utilities to prepare the data for machine-learning models, including scaling, feature extraction, and preprocessing. The relationship between these three is graphically depicted in Figure~\ref{fig:architecture}.
Supporting files like \texttt{LICENSE}, \texttt{README.md}, and \texttt{pyproject.toml} provide high level descriptions of functioning and installation, and straightforward to distribute via PyPI. The documentation is built using \texttt{Sphinx}~\cite{sphinx_2020_sphinx} from the source files contained in the \texttt{docs} folder.


\subsection{Scalar class}\label{subsubsec:Scalar_class}
The \texttt{Scalar} class was designed to efficiently manage and manipulate scalar fields defined over three-dimensional domains. Its core functionality is the data handling through file pointers, which enables working with multiple large matrices that, if loaded with commonly employed objects such as \texttt{numpy} arrays, would cause memory overloads. The class provides two operational modes: \textit{normal} and \textit{light mode}. The \textit{normal mode} assigns a numpy array to the object. On the other hand, when a path to a file is specified, the object defaults to \textit{light mode}, so that every time the object is called, it reads and returns the numerical values of the file located at the assigned path. While this provides better memory efficiency when working with a large number of files (e.g., with multiple timesteps or with many chemical species), as the files are only read when needed, the drawback of this approach is that the data must be reloaded from disk every time the object is accessed. This repeated I/O operation could potentially slow down the operations; however, as demonstrated in section \ref{subsec:Scalar_efficiency}, the load time remains relatively low---approximately one second for files containing over five hundred million grid points---making the approach suitable for many DNS dataset sizes. Along with this optimized access pattern, the class provides built-in support for extracting and plotting mid-plane slices along the x, y, and z directions, allowing users to quickly inspect cross-sectional features of the field. These plotting utilities are based on \texttt{Matplotlib}~\cite{Hunter2007matplotlib} and are designed for scientific data visualization. Multiple examples of the plots' appearance follow in section~\ref{sec:usage_examples}.

\subsection{Field class}\label{subsubsec:Field_class}
This class is a container of different \texttt{Scalar} objects, all associated with the same \texttt{Mesh}, which is a class that handles the mesh and related operations. The files are read from a folder that contains all the formatted data, the chemical mechanism used, the mesh, and an \texttt{info.json} file that contains the metadata. An example of a valid structured folder that can be read from the software is the following:

\dirtree{%
.1 .
.2 chem\_thermo\_tran.
.3 li\_h2.yaml.
.2 data.
.3 P\_Pa\_id000.dat.
.3 RHO\_kgm-3\_id000.dat.
.3 T\_K\_id000.dat.
.3 UX\_ms-1\_id000.dat.
.3 UY\_ms-1\_id000.dat.
.3 UZ\_ms-1\_id000.dat.
.3 YH2O2\_id000.dat.
.3 YH2O\_id000.dat.
.3 YH2\_id000.dat.
.3 YHO2\_id000.dat.
.3 YH\_id000.dat.
.3 YN2\_id000.dat.
.3 YO2\_id000.dat.
.3 YOH\_id000.dat.
.3 YO\_id000.dat.
.2 grid.
.3 X\_m.dat.
.3 Y\_m.dat.
.3 Z\_m.dat.
.2 info.json.
}
\vspace{8pt}
\noindent
where the \texttt{chem\_thermo\_tran} repository is only necessary for reacting flows DNS. For non-reacting flows, a trigger can be set so that the initialization method does not expect to find the chemical properties file.

The files in the \texttt{data} directory follow a structured naming convention that encodes the physical quantity, units, and a unique identifier. Each filename begins with an abbreviation of the measured quantity (e.g., \texttt{P} for pressure, \texttt{RHO} for density, \texttt{T} for temperature, \texttt{UX}, \texttt{UY}, \texttt{UZ} for velocity components, and \texttt{YH2O}, \texttt{YH2O2}, etc., for chemical species mass fractions). This is followed by the units in concise notation (e.g., \texttt{Pa} for pascals, \texttt{kgm-3} for kilograms per cubic meter, \texttt{K} for kelvin, \texttt{ms-1} for meters per second) and ends with \texttt{\_id000} as a unique identifier, used to represent different timeframes. The \texttt{.dat} extension indicates data files, stored in a 32-bit binary floating-point format. This formatting is coherent with the structure proposed in the BLASTNet repository, and ensures clear and organized file identification across simulations.

\subsubsection{Initialization and dynamic attributes}
The \texttt{\_\_init\_\_} method initializes the \texttt{Field} object by setting up directory paths, verifying the folder structure, loading configuration files, and building key attributes; the procedure evolves through the following steps:

\begin{enumerate}
\item \textit{Folder structure and files check}: the instantiation begins by checking that the specified folder follows the expected structure and contains the required data files.

\item \textit{Filter size extraction}: the method then identifies whether the field was filtered or not. This operation leverages the containing folder naming imposed by the filtering functions. If the field was previously filtered, the \texttt{\_\_init\_\_} method extracts the filter size, storing this value in a dedicated object attribute.

\item \textit{Configuration size load}: the \texttt{info.json} configuration file within the folder is read to set the dimensions of the 3D data field and assign it to the \texttt{self.shape} attribute.

\item \textit{Mesh attribute creation}: a \texttt{Mesh} object for spatial coordinates is constructed by reading the \texttt{X\_m.dat}, \texttt{Y\_m.dat}, and \texttt{Z\_m.dat} grid files. The current implementation supports structured grids only, reflecting the most widely adopted data format used by the reference databases targeted by the library. Support for unstructured or more complex mesh topologies is not precluded by the design and may be incorporated in future releases based on community needs.

\item \textit{Kinetic mechanism and species names extraction}: the kinetic mechanism file is loaded and species information is extracted; this step is avoided if the flag \texttt{reactive} is set to False.

\item \textit{Building scalar attributes}: finally, two lists of scalar attributes and paths are generated, followed by a call to the \texttt{update} method to initialize these attributes. This creates a dynamic attributes list that includes only the attributes with an associated file in the \texttt{data} directory whose names are formatted following the specifics of the module \texttt{variables.py}. 

\end{enumerate}

\noindent
Throughout the initialization, status updates are printed to the console, providing step-by-step feedback on the setup process.

\subsubsection{Filtering, downsampling and cutting}
Filtering, downsampling, and cutting operations are very common when DNS data are processed for \textit{a priori} model validation, and hence they were implemented in the library using one-line commands. These three methods share the general procedure, which consists of reading and processing sequentially all the files in the \texttt{data} and \texttt{grid} folders, saving them in a new container folder with an updated \texttt{info.json} file if the shape of the field changes during the operation.

For filtering operations, the filter kernel (box or Gaussian) must be specified, together with the possibility to perform Favre filtering. When using the Gaussian kernel, the filter is applied as a homogeneous Gaussian function, defined as
    \begin{equation}
        G(r) = \frac{1}{\sigma \sqrt{2 \pi}} \exp \left( -\frac{r^2}{2 \sigma^2} \right),    
    \end{equation}
    where \(\sigma\) is chosen such that the Gaussian filter has the same second moment as a box filter of width \(\delta\), satisfying
    \begin{equation}
        \sigma^2 = \frac{1}{12} \delta^2.
    \end{equation}

\noindent
This equivalence in second moments ensures that both the Gaussian and box filters provide comparable smoothing effects. The filter width \(\Delta\) is expressed in terms of multiples of the DNS grid size:
\begin{equation}
    \Delta = \frac{\delta}{\delta_{\text{DNS}}},
\end{equation}
where \(\delta\) represents the desired filter width and \(\delta_{\text{DNS}}\) is the DNS grid spacing.

\subsubsection{Gradients computation}
The computation of gradients is consistent with any filtering operations applied to the field. Specifically, when the field is both filtered and downsampled, gradients can be computed on adjacent cells, as the resulting field resembles a coarse-grained solution typically available in Large Eddy Simulation (LES) software. When the field is filtered but not downsampled, gradient computation is performed on every possible virtually downsampled grid by effectively skipping \( \Delta \) adjacent cells, where \( \Delta \) is relative to the filter size.

The current implementation employs high-order central difference schemes for spatial derivatives based on \cite{findiff}, which allows arbitrarily high order schemes. \texttt{aPriori} defaults to fourth-order accuracy for both the gradient and the Laplacian. When handling filtered fields, the derivative order defaults to second order, as many LES solvers---particularly those operating on unstructured grids---typically rely on lower-order discretizations. The default accuracy can be easily modified using the \texttt{set\_gradients\_order} function, without requiring changes to the source code.

Nevertheless, in cases where specialized numerical schemes are required for accurate post-processing—for example, to mimic the discretization used by the DNS solver that generated the data—users may modify the gradient schemes directly. This can be achieved by overriding the external gradient and Laplacian functions defined in the \texttt{derivatives.py} module. These functions are invoked by all gradient-related methods within the \texttt{Field} class. Overriding them ensures consistent gradient computation throughout the library, providing both flexibility and numerical consistency in advanced scenarios.

\subsubsection{Data Chunking}

While handling large variables via pointers can be efficient for operations involving only a small number of scalars---ideally one scalar at a time---chemistry-related computations typically require simultaneous access to the full set of species concentrations and temperature. To address memory constraints in such cases, data chunking is implemented.
This approach is particularly important for operations involving the full thermochemical state vector, such as reaction rate computations (which are almost always required during simulations, since source terms are generally not stored to conserve disk space) as well as CSP analyses and local reactor-based evaluations.

In practice, the files representing the thermochemical state vector are read in discrete chunks: only a subset of the data is loaded into memory at a time, processed, and then written to the corresponding output files. These output files remain open throughout the procedure to accommodate results from successive chunks. All utilities that perform such operations include an option to specify the number of chunks to be used; by default, this value is set to 1000.


\section{Usage examples}\label{sec:usage_examples}
\subsection{Scalar class memory usage}\label{subsec:Scalar_efficiency}

This example demonstrates how the \texttt{Scalar} class can be used to efficiently manage large DNS-like datasets while minimizing RAM usage. The benchmark creates a set of 3D scalar fields with increasingly growing sizes, stores them on disk, and then uses the \texttt{Scalar} class in light mode to measure the time required to access each file. The results are plotted as load time versus memory size.

The script begins by importing required libraries for data handling, visualization, and benchmarking. 

\begin{lstlisting}[style=mystyle, language=Python]
import aPriori as ap
import numpy as np
import matplotlib.pyplot as plt
import os
import sys
import time
\end{lstlisting}

\noindent
The following lines initialize a list of 3D shapes, starting from a base size of $40 \times 30 \times 15$, and iteratively scale each dimension by a factor of 1.1 over 39 steps. This results in a sequence of 40 shapes with increasing total number of grid points, ranging from $1.8 \times 10^4$ up to roughly $5.4 \times 10^8$. For each shape, a scalar field filled with random floating-point values is generated using NumPy, and saved to disk in binary format using the \texttt{save\_file()} utility. This step generates output files which resemble BLASTNet formatting.

\begin{lstlisting}[style=mystyle, language=Python]
s0 = [40, 30, 15]
f = 1.1
n = 39
shapes = [s0]

for _ in range(n):
    next_size = (np.array(shapes[-1]) * f).astype(int).tolist()
    shapes.append(next_size)

elements = [np.prod(shape) for shape in shapes]

data_folder = 'data'
if not os.path.exists(data_folder):
    os.mkdir(data_folder)

filenames = [os.path.join(data_folder, 'f_{}_{}_{}.dat'.format(*shape)) for shape in shapes]
scalars = [np.float32(np.random.rand(*shape)) for shape in shapes]

for scalar, filename in zip(scalars, filenames):
    ap.save_file(scalar, file_name=filename)
\end{lstlisting}

\noindent
Subsequently we instantiate \texttt{Scalar} objects by pointing to the saved files, without loading them into memory. This allows to measure the overhead associated with delayed loading.

\begin{lstlisting}[style=mystyle, language=Python]
ram_size = [scalar.nbytes / 1024**2 for scalar in scalars]
light_scalars = [ap.Scalar(shape=shape, path=filename) for shape, filename in zip(shapes, filenames)]
disk_sizes = [os.path.getsize(filename) / 1024**2 for filename in filenames]
ram_size_light = [sys.getsizeof(light_scalar) for light_scalar in light_scalars]
\end{lstlisting}

\noindent
We access each scalar field using the \texttt{.value} property and record the time required to read the data from disk.

\begin{lstlisting}[style=mystyle, language=Python]
t_load = []
for s in light_scalars:
    t0 = time.time()
    array = s.value  # loading the array
    t1 = time.time()
    t_load.append(t1 - t0)
\end{lstlisting}

\begin{figure}[h!]
    \centering
    \includegraphics[width=0.95\linewidth]{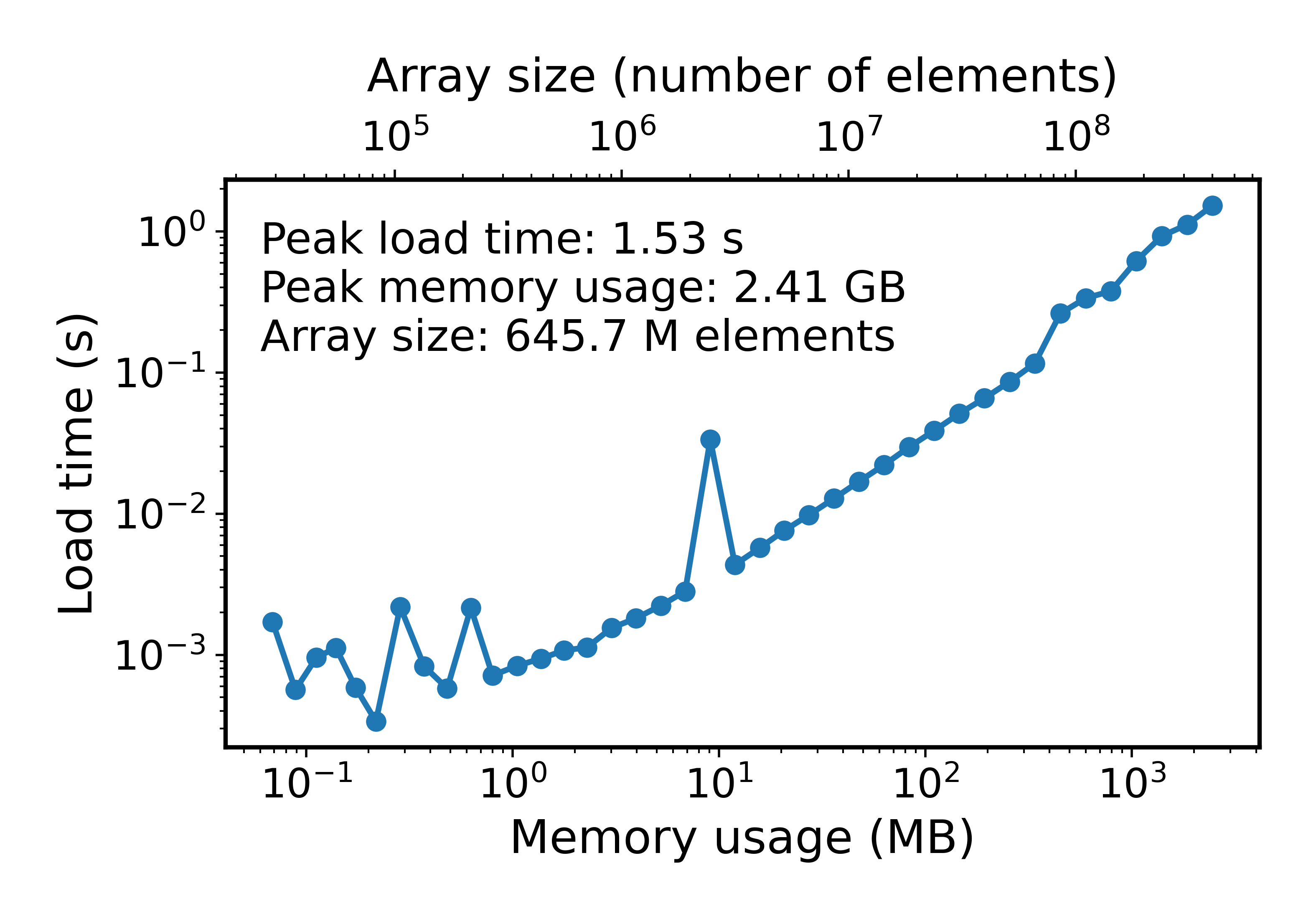}
    \caption{
    Load time of the \texttt{Scalar} class operating in light mode, measured by reading binary DNS-like scalar fields of increasing size from disk. 
    Each point corresponds to a single array whose dimensions were progressively scaled, resulting in memory footprints ranging from a few hundred kilobytes to several gigabytes. 
    The logarithmic axes reveal a near-linear growth of read time with respect to array size, with the largest field tested (approximately $6.46\times10^8$ elements, 2.41\,GB) requiring only 1.53\,s to load. 
    This benchmark illustrates the efficiency of the pointer-based data access strategy used in the software, enabling manipulation of very large DNS fields without exhausting the system's memory.}
    \label{fig:pointer_efficiency}
\end{figure}

\noindent
Finally, a logarithmic plot is displayed in Figure~\ref{fig:pointer_efficiency} to show how the load time scales with the field size, together with a note which underlines the peak memory and associated reading time.

\begin{lstlisting}[style=mystyle, language=Python]
# Create figure and primary axis
fig, ax1 = plt.subplots(figsize=(6.3, 4.4), dpi=500)

# Plot on log-log scale
ax1.set_xscale('log')
ax1.set_yscale('log')
ax1.plot(ram_size, t_load, marker='o', linestyle='-', linewidth=2, markersize=6, color='C0')
ax1.set_xlabel('Memory usage (MB)', fontsize=16)
ax1.set_ylabel('Load time (s)', fontsize=16)
ax1.tick_params(axis='both', labelsize=14)
# ax1.grid(True, which='both', linestyle='--', alpha=0.2)
ax1.grid(False)

# Mapping functions for secondary x-axis
def ram_to_size(x):
    return np.interp(x, ram_size, elements)

def size_to_ram(x):
    return np.interp(x, elements, ram_size)

# Add log-scale secondary x-axis
ax2 = ax1.secondary_xaxis('top', functions=(ram_to_size, size_to_ram))
ax2.set_xscale('log')
ax2.set_xlabel('DNS grid points', fontsize=16, labelpad=10)
ax2.tick_params(axis='x', labelsize=14)

# Get max values
max_time = np.max(t_load)
max_ram = ram_size[np.array(t_load).argmax()]  # RAM at max load time
max_elems = elements[-1]

# Create annotation text
annotation_text = (
    f"Peak load time: {max_time:.2f} s\n"
    f"Peak memory usage: {max_ram / 1024:.2f} GB\n"
    f"Array size: {max_elems / 1e6:.1f} M elements"
)

# Add to plot (adjust x and y coordinates as needed)
plt.text(0.06, 1.2, annotation_text,
        fontsize=15, verticalalignment='top',
        #bbox=dict(boxstyle="round,pad=0.4", facecolor="white", edgecolor="gray", alpha=0.9)
        )

# Final layout
fig.tight_layout()

plt.savefig('Memory_load.png')
plt.show()
\end{lstlisting}

\subsection{Data reading and visualization}

This section demonstrates how the \texttt{aPriori} package can be used to load and visualize DNS datasets from structured folders. Two flame configurations are processed: a slot turbulent lifted hydrogen flame in a hot coflow \cite{Yoo2009h2lifted, Jung2021} and a turbulent jet premixed methane/air flame \cite{Brouzet_Talei_Brear_Cuenot_2021}.

We begin by importing the required modules and instantiating a \texttt{Field} object that points to the folder containing the DNS data of the hydrogen flame. Scalar fields are accessed via their variable names, and visualized along the $z$ mid-plane using built-in plotting utilities. Orientation and title control options are provided for publication formatting.

\begin{lstlisting}[style=mystyle, language=Python]
import aPriori as ap
import matplotlib.pyplot as plt

h2_flame = ap.Field('DNS_DATA_2d')

f1, ax1 = h2_flame.plot_z_midplane('T', transpose=True, remove_title=True)
f2, ax2 = h2_flame.plot_z_midplane('U_X', transpose=True, log=False, remove_title=True)
\end{lstlisting}

\noindent
We compute the strain rate tensor using a method of the \texttt{Field} class. The resulting scalar with the strain rate magnitude is automatically added to the object as a dynamic attribute, and is then visualized. Finally, we plot the mass fraction of \ce{H2O}, specifying threshold values to highlight regions of interest.

\begin{lstlisting}[style=mystyle, language=Python]
h2_flame.compute_strain_rate(save_tensor=False)

f3, ax3 = h2_flame.plot_z_midplane('S_DNS', transpose=True, log=True, remove_title=True)
f4, ax4 = h2_flame.plot_z_midplane(
    'YH2O', 
    transpose=True, 
    vmin=0.03, 
    levels=[0.03, 0.10, 0.15],
   remove_title=True)
\end{lstlisting}

\noindent
Each subplot is annotated using the \texttt{text()} function to add panel labels and variable units. Figures are then saved with transparent backgrounds and tight bounding boxes.

\begin{lstlisting}[style=mystyle, language=Python]
ax1.text(0.0, 1.17, 'a', fontsize=55, fontweight='bold', fontname='arial', transform=ax1.transAxes)
ax1.text(1.15, 1.0, 'T [K]', fontsize=36, transform=ax1.transAxes)
f1.savefig('figures/fig_a.png', transparent=True, bbox_inches='tight')

ax2.text(0.0, 1.17, 'b', fontsize=55, fontweight='bold', transform=ax2.transAxes)
ax2.text(1.15, 1.0, r'$U_x$ [m/s]', fontsize=36, transform=ax2.transAxes)
f2.savefig('figures/fig_b.png', transparent=True, bbox_inches='tight')

ax3.text(0.0, 1.17, 'c', fontsize=55, fontweight='bold', transform=ax3.transAxes)
ax3.text(1.15, 1.0, r'S [$s^{-1}$]', fontsize=36, transform=ax3.transAxes)
f3.savefig('figures/fig_c.png', transparent=True, bbox_inches='tight')

ax4.text(0.0, 1.17, 'd', fontsize=55, fontweight='bold', transform=ax4.transAxes)
ax4.text(1.15, 1.0, r'$Y_{H2O}$ [-]', fontsize=36, transform=ax4.transAxes)
f4.savefig('figures/fig_d.png', transparent=True, bbox_inches='tight')
\end{lstlisting}

\noindent
Next, we apply the same procedure to a CH\textsubscript{4} premixed flame. After loading the cut-domain dataset, we compute a progress variable based on O\textsubscript{2} and visualize it with a custom colormap.

\begin{lstlisting}[style=mystyle, language=Python]
ch4_flame = ap.Field('CH4_flame_cut')
ch4_flame.compute_progress_variable('O2')

f5, ax5 = ch4_flame.plot_z_midplane('C', transpose=True, colormap='inferno',
                                    vmin=0, vmax=1, remove_title=True)
\end{lstlisting}

\noindent
We then compute the velocity magnitude and the gradient of the progress variable, and visualize both on the same mid-plane. The use of reversed colormaps and logarithmic scaling improves visual contrast for fields spanning multiple orders of magnitude.

\begin{lstlisting}[style=mystyle, language=Python]
ch4_flame.compute_velocity_module()
f6, ax6 = ch4_flame.plot_z_midplane('U_DNS', transpose=True, colormap='inferno_r',
                                    remove_title=True)

ch4_flame.compute_gradient_C()
f7, ax7 = ch4_flame.plot_z_midplane('C_grad', transpose=True, colormap='inferno_r',
                                    log=True, vmin=1.5, transparent=True,
                                    remove_title=True)

f8, ax8 = ch4_flame.plot_z_midplane('YCH2O', transpose=True, colormap='inferno',
                                    log=True, vmin=-5, remove_title=True)
\end{lstlisting}

\begin{figure*}[ht!]
    \centering
\includegraphics[width=1.0\linewidth,trim=5 5 5 5,clip]{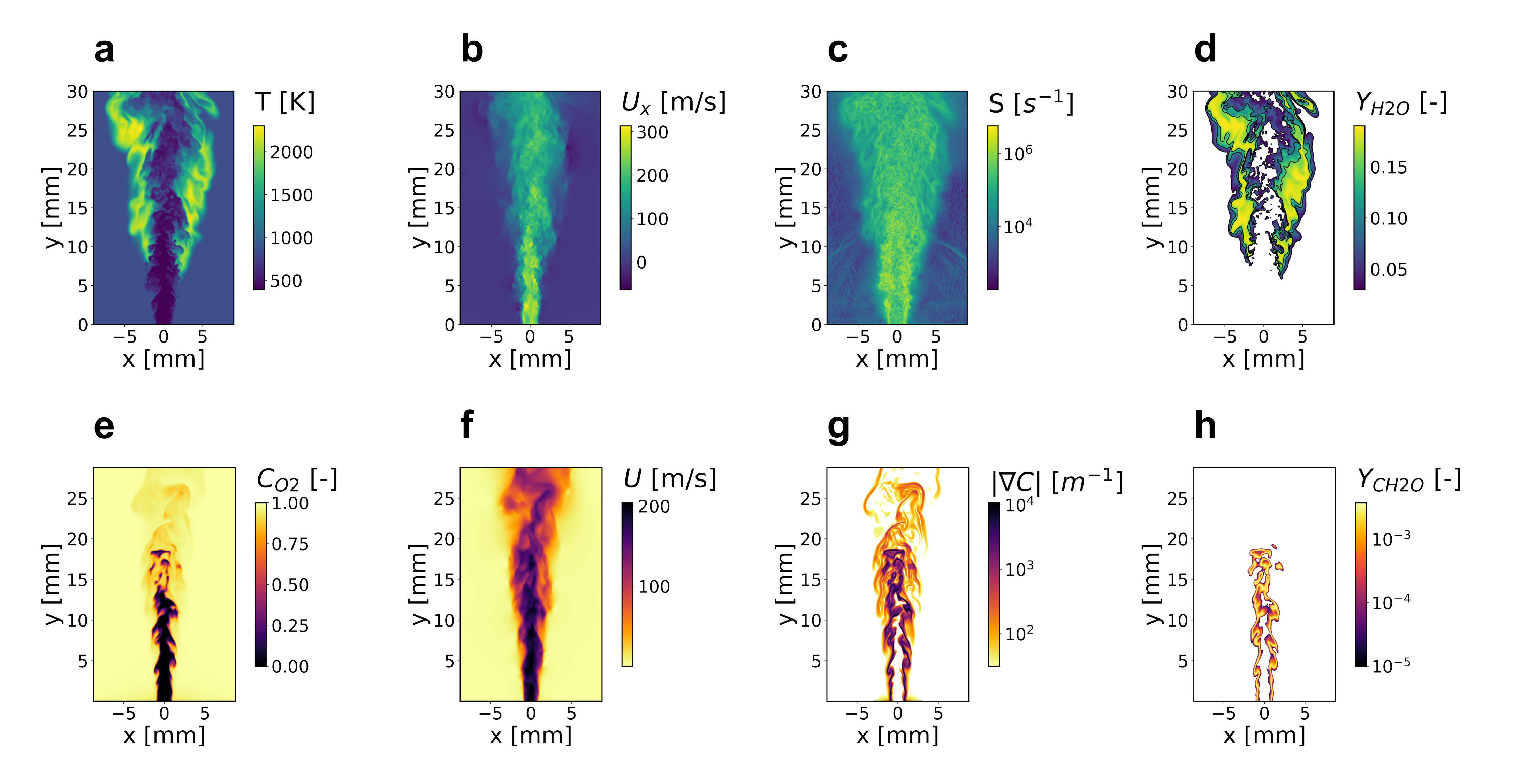}
    \caption{
    DNS mid-plane visualisations produced with the \texttt{aPriori} plotting utilities for two turbulent flame configurations. 
    (a)--(d) Lifted hydrogen flame in a heated coflow: temperature, streamwise velocity $U_x$, strain-rate magnitude $\mathcal{S} = (2 S_{ij} S_{ij})^{1/2}$ on a logarithmic scale, and water vapour mass fraction $Y_{\mathrm{H_2O}}$. 
    (e)--(h) Premixed methane/air flame: progress variable based on $\mathrm{CO_2}$, velocity magnitude $U$, progress-variable gradient $\lvert \nabla C \rvert$ (logarithmic scale), and formaldehyde mass fraction $Y_{\mathrm{CH_2O}}$.}
    \label{fig:plotting_utils}
\end{figure*}

\subsection{Filtering and downsampling}
In this section an example code illustrates the filtering, downsampling, and subdomain extraction utilities. The dataset used is a non-reacting, forced, homogeneous isotropic turbulence DNS, from a previous work of Gauding et al. \cite{Gauding_2022}.

The first step consists of instantiating the field object, using as a reference the data containing folder. As the case is non-reactive, a corresponding flag must be set accordingly to avoid the initialization process looking for the chemical properties folder. The velocity module is then computed and visualized.

\begin{lstlisting}[style=mystyle, language=Python]
import pathlib
import pandas as pd
import numpy as np
import matplotlib.pyplot as plt
import aPriori as ap
import os

data_folder = os.path.join('DNS_turbulent_transport', 'Filtering_example')

# Initialize the field, with the option 'reactive' set to false
DNS_field = ap.Field(data_folder, reactive=False)

DNS_field.compute_velocity_module()
f, ax = DNS_field.plot_z_midplane(
    'U_DNS', 
    scale='m', 
    colormap='Blues',
    vmin=0,
    vmax=11,
    cbar_title=r'U [m/s]',
    y_ticks=[2, 4, 6],
    remove_title=True
    )

\end{lstlisting}

\noindent
Every time a field is filtered, cut, or downsampled, the resulting output is saved in a replicated folder, which has the same structure of the original one. This allows the library to read the resulting output field as a different object. At this stage the pointer-based structure of the code comes in useful, as it allows to store all these objects in a dictionary and work with them simultaneously. 

\noindent
In the following, three different filter sizes ($\Delta = $ 8, 16, and 32) are selected. The \texttt{filter()} method of the class performs the filtering operations, creates the filtered field folder, and returns the folder's name as a string. The folder name contains information regarding the filter size, the filter kernel, and whether Favre filtering was employed. This string is then used to instantiate the object as a new instance of the class \texttt{Field}, which is then added to the dictionary with a proper key representing the filter size.

\begin{lstlisting}[style=mystyle, language=Python]
filter_sizes = [8, 16, 32]
filtered_fields = dict()

for delta in filter_sizes:
    f_name = DNS_field.filter(delta)
    f = ap.Field(f_name, reactive=False)
    filtered_fields[f'{delta}'] = f
    del f
\end{lstlisting}

\noindent
The dictionary is then browsed and the velocity magnitude values are accessed to be plotted for all the filter sizes.

\begin{lstlisting}[style=mystyle, language=Python]
figures_dir = os.path.join(data_folder,'Figures')
os.mkdir(figures_dir)
for i, delta in enumerate(filter_sizes):
    # apply different figure properties depending on the position
    if i == 0:
        remove_y = False
    else:
        remove_y = True
    if i == len(filter_sizes)-1:
        remove_cbar = False
        cbar_title = r'U [m/s]'
    else:
        remove_cbar = True
        cbar_title = None
        
    # Plot and save figure and axis objects
    f, ax = filtered_fields[f'{delta}'].plot_z_midplane('U_DNS', scale='m', colormap='Blues', title=r'$\Delta$'+f' = {delta}', vmin=0, vmax=11, cbar_title=cbar_title, y_ticks=[2, 4, 6], remove_x=True, remove_y=remove_y, remove_cbar=remove_cbar)
    
    # Save figure
    f.savefig(os.path.join(figures_dir,f'delta{delta}.png'), 
              transparent=True, 
              bbox_inches='tight'
              )
\end{lstlisting}    

\noindent
The same procedure is here applied to downsample and plot the filtered fields.

\begin{lstlisting}[style=mystyle, language=Python]
downsampled_fields = dict() # initialize an empty dictionary
for delta in filter_sizes:
    # Downsample the filtered field and initialize a new one using the new directory
    ds = ap.Field(filtered_fields[f'{delta}'].downsample(ds_size=delta), reactive=False) 
    downsampled_fields[f"{delta}"] = ds
    del ds

for i, delta in enumerate(filter_sizes):
    if i == 0:
        remove_y = False
    else:
        remove_y = True
    if i == len(filter_sizes)-1:
        remove_cbar = False
        cbar_title = r'U [m/s]'
    else:
        remove_cbar = True
        cbar_title = None
        
    f, ax = downsampled_fields[f"{delta}"].plot_z_midplane('U_DNS', scale='m', colormap='Blues', title=r'$\Delta$'+f' = {delta}', vmin=0, vmax=11, cbar_title=cbar_title, remove_title=True, y_ticks=[2, 4, 6], remove_x=False, remove_y=remove_y, remove_cbar=remove_cbar)
    
    f.savefig(os.path.join(figures_dir,f'delta{delta}DS.png'), 
              transparent=True, 
              bbox_inches='tight'
              )
\end{lstlisting}

Figure \ref{fig:plotting_utils} shows the outputs of the code presented in this section.

\begin{figure}[ht!]
    \centering
\includegraphics[width=1.0\linewidth,trim=5 5 5 5,clip]{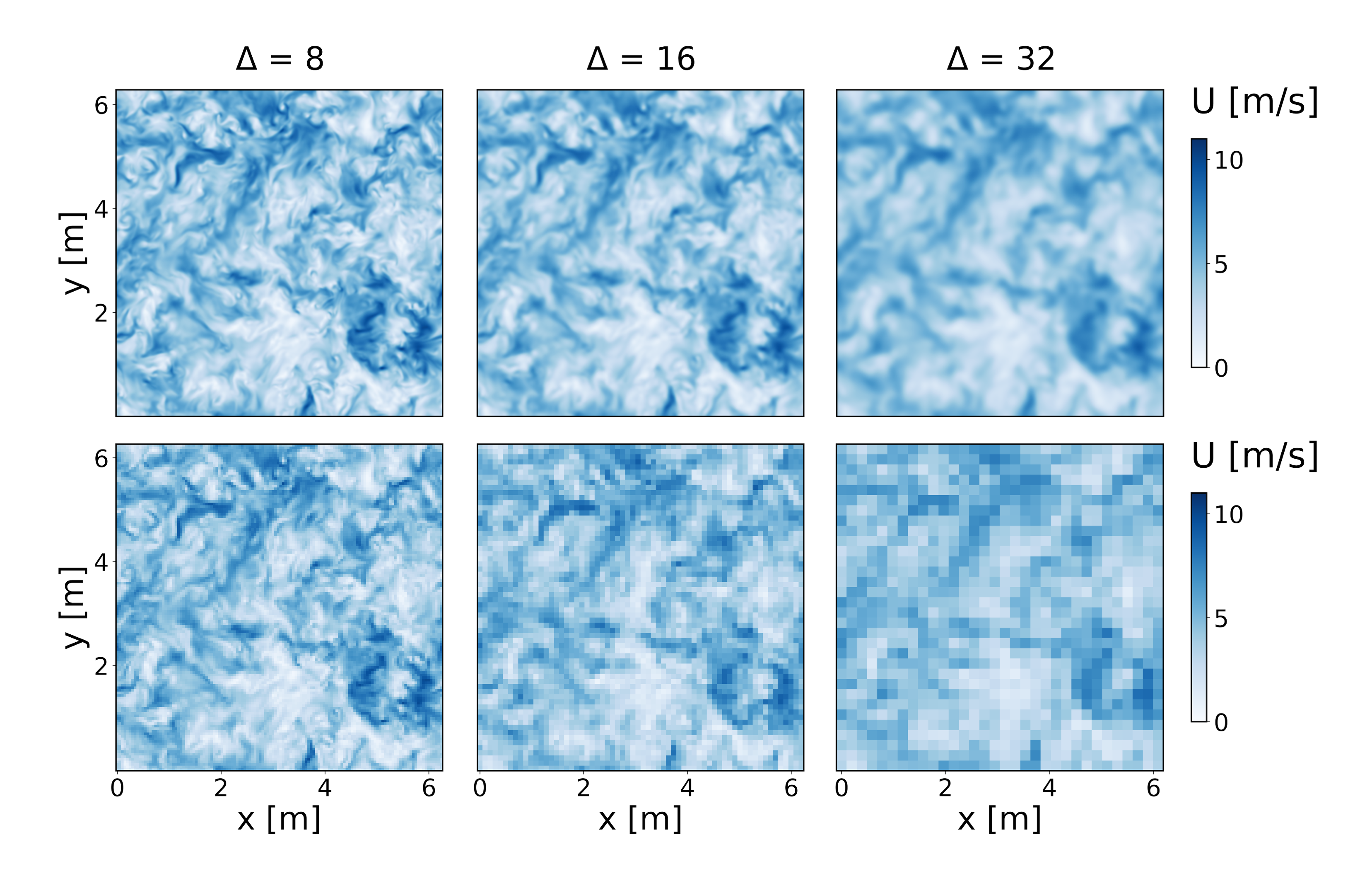}
    \caption{Mid-plane visualization of the velocity field from the homogeneous isotropic turbulence DNS of Gauding et al. \cite{Gauding_2022}. Each column corresponds to a different filter size. The top row shows the filtered velocity magnitude, while the bottom row presents the same filtered fields after downsampling.}
    \label{fig:plotting_utils}
\end{figure}

\subsection{Turbulence model assessment}
Model validation is one of the key functionalities that the library was designed for. In section~\ref{sec:general-framework-apriori} we first define the general framework used for \textit{a priori} model validation, as it was fundamental in defining various aspects of the software architecture, and in section~\ref{sec:smagorinsky-example} we show an example application.

\subsubsection{General framework for \textit{a priori} validation} \label{sec:general-framework-apriori}
The framework to perform \textit{a priori} validation of sub-filter models with the library consists roughly in the following steps:
\begin{enumerate}
    \item The derived quantities of interest are computed from the original DNS field, relying on the variables which constitute the solution of the partial differential equations.
    \item The DNS field is filtered. As shown in the previous example, the filtering operation will create an additional folder that contains the filtered data. The main variables (e.g., velocity, density, etc...) constitute the \textit{LES-like} \footnote{The difference between a filtered DNS and an LES simulation, which can be reduced to the discrepancy between the filtered solution and the solution of the filtered equations, is a widely discussed topic in the literature, for example by Pope \cite{Pope2004-gv}, and will not be further developed here. Still it should be noted that this discrepancy constitutes a theoretical limitation of the \textit{a priori} validation approach based on DNS data.} filtered field, meaning that those quantities resemble what is ideally accessible in a LES simulation \footnote{In the present article, we consider the \textit{LES-like} values equivalent to the  solution of the filtered equations.}. The non-linear quantities computed on the DNS at step one, such as the kinetic energy, or the strain rate, are filtered as well from this operation, and they represent variables containing sub-filter information, which is typically not accessible in LES.
    \item The derived quantities of interest are computed from the filtered field. For example, the x-component of velocity, in the software, is referred to as \texttt{U\_X} both in the unfiltered and in the filtered field. In the unfiltered field it represents the DNS velocity $U_x$, in the filtered one it represents $\overline{U}_x$. These two quantities can be accessed with the following commands, respectively:
    \begin{lstlisting}[style=mystyle, language=Python]
DNS_field.U_X
filtered_field.U_X\end{lstlisting}
    given that \texttt{DNS\_field} and the object containting the DNS data and \texttt{filtered\_field} is the object containing the field filtered from the aforementioned one.
\end{enumerate}
To give an example with the kinetic energy, the procedure would look as follows:
\begin{enumerate}
    \item Compute the kinetic energy $k_{DNS}$:
    \begin{equation}
        k_{DNS} \equiv \frac{1}{2}\mathbf{U}\cdot \mathbf{U};
    \end{equation}
    \item Filter the DNS field: additionally to the main variables, one obtains the filtered kinetic energy, which reads
    \begin{equation}
        \overline{k}_{DNS} \equiv \overline{\frac{1}{2}\mathbf{U}\cdot \mathbf{U}};
    \end{equation}
    \item From the \textit{LES-like} field, compute again the resolved kinetic energy:
    \begin{equation}
        \overline{k}_{LES} \equiv \frac{1}{2} \overline{\mathbf{U}} \cdot \overline{\mathbf{U}};
    \end{equation}
    now the residual kinetic energy can be computed as
    \begin{equation}
        k_r =  \overline{\frac{1}{2}\mathbf{U}\cdot \mathbf{U}} - \frac{1}{2} \overline{\mathbf{U}} \cdot \overline{\mathbf{U}} \equiv \overline{k}_{DNS} - \overline{k}_{LES}
    \end{equation}
    with $k_r$, $\overline{k}_{DNS}$ and $\overline{k}_{LES}$ being attributes of the filtered field object.
\end{enumerate}

\subsubsection{Smagorinsky model} \label{sec:smagorinsky-example}
The following example shows how to compute sub-filter quantities related to turbulence, specifically the residual dissipation rate and the residual stresses.

The code starts by importing the libraries and declaring the input variables:

\begin{lstlisting}[style=mystyle, language=Python]
import numpy as np
import aPriori as ap
import os

#============ Input variables ============#
data_folder = 'DNS_turbulent_transport'
filter_size = 8
Cs = 0.16
figures_folder = 'figures_subgrid_turbulence'
#=========================================#
\end{lstlisting}

\noindent
The following steps consist in creating a folder to save the figures, initializing the unfiltered DNS field object, and leveraging the built-in methods to compute the rate-of-strain tensor, which is defined as
\begin{equation}
    S_{ij}^{DNS} \equiv \frac{1}{2} \left( \frac{\partial U_i}{\partial x_j} + \frac{\partial U_j}{\partial x_i} \right),
\end{equation}
with $\mathcal{S}$ representing its module:
\begin{equation}
    \mathcal{S}_{DNS} \equiv \left( 2S_{ij} S_{ij} \right)^{1/2}
\end{equation}

\begin{lstlisting}[style=mystyle, language=Python]
# Create the folder for the output figures
if not os.path.exists(figures_folder):
    os.mkdir(figures_folder)

# Initialize the field, with the option 'reactive' set to false
DNS_field = ap.Field(os.path.join(data_folder, 'DNS_data_cut'), reactive=False)

# DNS Strain rate computation
DNS_field.compute_strain_rate(save_derivatives=False, save_tensor=True)

# Strain rate module plot
f1,ax1 = DNS_field.plot_z_midplane(
    'S_DNS', 
    log=True, 
    vmax=2.6, 
    vmin=.5, 
    colormap='Blues',
    scale='m',
    title=r'$\mathcal{S}=2S_{ij}S_{ij}$',
    remove_cbar=True,
    transparent=False,
    )
f1.savefig(
    os.path.join(figures_folder, 'f1'), 
    transparent=True,
    bbox_inches='tight')
\end{lstlisting}

\noindent
The DNS field is then filtered with a filter size $\Delta$, specified in the code inputs. The filtering function returns a string that is used to initialize the filtered field object, which contains the filtered rate of strain
\begin{equation}\label{eq:strain_rate_DNS_filtered}
    \overline{S}_{ij}^{DNS} \equiv \overline {\frac{1}{2} \left( \frac{\partial U_i}{\partial x_j} + \frac{\partial U_j}{\partial x_i} \right)}.
\end{equation}

\begin{lstlisting}[style=mystyle, language=Python]
# Filtering field.
# The function will return the relative path of the filtered field folder
f_name = DNS_field.filter(filter_size=filter_size, filter_type='Gauss')
# The filtered field can now be initialized
filtered_field = ap.Field(f_name, reactive=False)

# DNS filtered dissipation rate plot
f2, ax2 = filtered_field.plot_z_midplane(
    'S_DNS', 
    log=True, 
    vmax=2.6, 
    vmin=.5, 
    colormap='Blues',
    scale='m',
    title=r'$\overline{\mathcal{S}}_{DNS} \equiv \overline{2S_{ij}{S}_{ij}}$',
    remove_y=True,
    remove_cbar=True,
    transparent=False,
    )
f2.savefig(
    os.path.join(figures_folder, 'f2'), 
    transparent=True,
    bbox_inches='tight')
\end{lstlisting}

\noindent
The strain rate can then be computed from the \textit{LES-like} field, leading to the following expression:
\begin{equation} \label{eq:strain_rate_LES}
    \overline{S}_{ij}^{LES} \equiv  \frac{1}{2} \left( \frac{\partial \overline{U}_i}{\partial x_j} + \frac{\partial \overline{U}_j}{\partial x_i} \right).
\end{equation}
The expressions in eq. \ref{eq:strain_rate_DNS_filtered} and eq. \ref{eq:strain_rate_LES} are then compared in a parity plot.
\begin{lstlisting}[style=mystyle, language=Python]
# LES Strain rate plot
filtered_field.compute_strain_rate(save_derivatives=False, save_tensor=True)
f3, ax3 = filtered_field.plot_z_midplane(
    'S_LES', 
    log=True, 
    vmax=2.6, 
    vmin=.5, 
    colormap='Blues',
    scale='m',
    title=r'$\overline{\mathcal{S}}_{LES} \equiv 2\overline{S}_{ij}\overline{S}_{ij}$',
    remove_y=True,
    transparent=False,
    )
f3.savefig(os.path.join(figures_folder, 'f3'), 
           transparent=True,
           bbox_inches='tight')

f4, ax4 = ap.parity_plot(
    filtered_field.S_DNS.value, 
    filtered_field.S_LES.value, 
    density=True,
    cmin=1e-9,
    ticks=[100, 200, 300, 400],
    rel_error=None,
    x_name=r'$\overline{\mathcal{S}}_{DNS}$',
    y_name=r'$\bar{\mathcal{S}}_{LES}$',
    )
f4.savefig(
    os.path.join(figures_folder, 'f4'), 
    transparent=True,
    bbox_inches='tight')
\end{lstlisting}

\begin{figure*}[ht!]
    \centering
\includegraphics[width=1.0\linewidth,trim=10 10 10 10,clip]{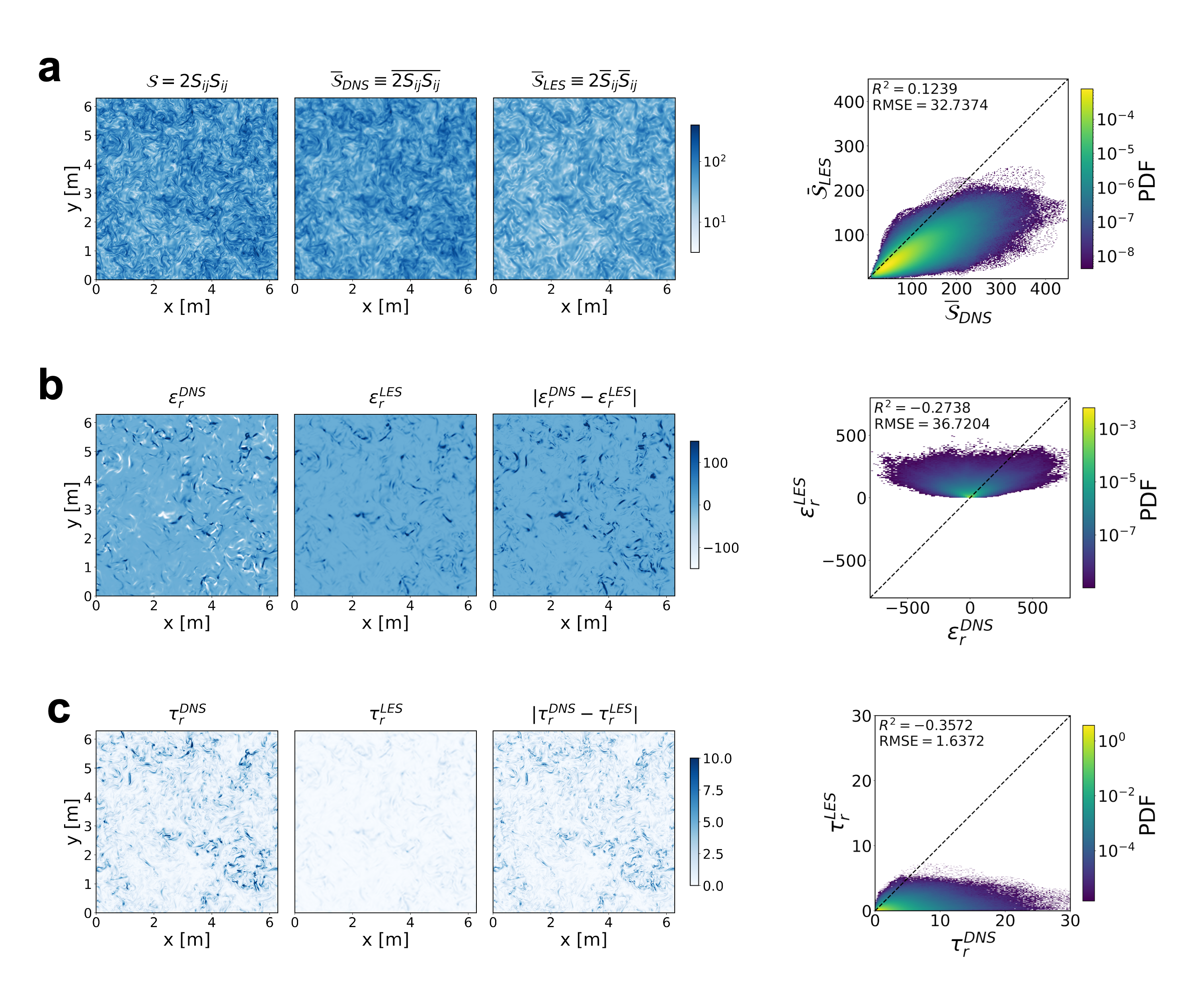}
    \caption{
    Assessment of subgrid‐scale quantities relevant to turbulence modelling using the \texttt{aPriori} framework. Data from~\cite{Gauding_2022}. 
    (a) Strain–rate magnitude $\mathcal{S}$ from DNS, its filtered counterpart $\overline{\mathcal{S}}_{\mathrm{DNS}}$, and the strain–rate magnitude computed from the filtered velocity field $\overline{\mathcal{S}}_{\mathrm{LES}}$, together with the corresponding parity plot. 
    (b) Residual dissipation rate $\epsilon_r^{\mathrm{DNS}}$ obtained from DNS, the Smagorinsky prediction $\epsilon_r^{\mathrm{LES}}$, the absolute error field, and the associated parity plot. 
    (c) Deviatoric residual stress module $\tau_r^{\mathrm{DNS}}$, its Smagorinsky estimate $\tau_r^{\mathrm{LES}}$, the absolute error field, and the corresponding parity plot. 
    All results refer to a homogeneous isotropic turbulence DNS filtered with $\Delta = 8$, illustrating the workflow for \textit{a priori} evaluation of subgrid‐scale closures.}
    \label{fig:smagorinsky_model}
\end{figure*}

\noindent
The results obtained with the code provided in the first part of this section are displayed in Figure \ref{fig:smagorinsky_model}.a.

We proceed computing quantities of interest for modeling, introducing Smagorinsky model and validating it against the DNS benchmark values. This constitutes an example of how the library can be used to validate existing closure models.
The transfer of energy from the residual scales to the filtered ones appears in the filtered energy equation and according to \cite{Pope2012-jn} reads
\begin{equation}\label{eq:residual_dissipation_rate}
    \epsilon_r \equiv -\tau_{ij}^r \overline{S}_{ij}^{LES},
\end{equation}
with $\tau_{ij}^r$ being the deviatoric part of the residual stress
\begin{equation}
    \tau_{ij}^r = \tau_{ij}^R - \frac{1}{3} \delta_{ij} \tau_{kk}^R ~,
\end{equation}
and $\tau_r^{\mathrm{DNS}}$ the magnitude of such tensor:
\begin{equation}
\tau_r^{\mathrm{DNS}} \equiv \left( 2 \tau_{ij}^r \tau_{ij}^r \right)^{1/2}.
\end{equation}

\begin{lstlisting}[style=mystyle, language=Python]
# Compute residual dissipation rate
filtered_field.DNS_folder_path = DNS_field.folder_path
filtered_field.compute_residual_dissipation_rate(mode='DNS')
filtered_field.Cs=Cs # set the Smagorinsky constant
filtered_field.compute_residual_dissipation_rate(mode='Smag')

# Compute absolute error, save it to file, and add it as a variable
absolute_error = np.abs(filtered_field.Epsilon_r_DNS.value - filtered_field.Epsilon_r_Smag.value)
error_file_name = 'Epsilon_r_AE_id000.dat'
error_file_path = os.path.join(filtered_field.data_path, error_file_name)
ap.save_file(absolute_error, error_file_path)
ap.add_variable('Epsilon_r_AE', 'Epsilon_r_AE_{}.dat')
filtered_field.update(verbose=True)

# Residual dissipation rate plot
f5, ax5 = filtered_field.plot_z_midplane(
    'Epsilon_r_DNS',  
    vmin=-150, 
    vmax=150, 
    transparent=False,
    title=r'$\epsilon_r^{DNS}$',
    scale='m',
    remove_cbar=True,
    colormap='Blues'
    )
f5.savefig(
    os.path.join(figures_folder, 'f5'), 
    transparent=True,
    bbox_inches='tight')

# Smagorinsky modeled residual dissipation rate
f6, ax6 = filtered_field.plot_z_midplane(
    'Epsilon_r_Smag', 
    vmin=-150, 
    vmax=150, 
    title=r'$\epsilon_r^{LES}$',
    transparent=False,
    scale='m',
    remove_y=True,
    remove_cbar=True,
    colormap='Blues'
    )
f6.savefig(
    os.path.join(figures_folder, 'f6'), 
    transparent=True,
    bbox_inches='tight')

f7,ax7 = filtered_field.plot_z_midplane(
    'Epsilon_r_AE', 
    scale='m',
    colormap='Blues',
    title=r'$|\epsilon_r^{DNS} - \epsilon_r^{LES}|$',
    vmin=-150,
    vmax=150,
    transparent=False,
    remove_y=True
    )
f7.savefig(
    os.path.join(figures_folder, 'f7'), 
    transparent=True,
    bbox_inches='tight')

# Residual dissipation rate parity plot
f8, ax8 = ap.parity_plot(
    filtered_field.Epsilon_r_DNS.value, 
    filtered_field.Epsilon_r_Smag.value, 
    density=True,
    cmin=1e-9,
    limits=[-800, 800],
    ticks=[-500, 0, 500],
    rel_error=None,
    x_name=r'$\epsilon_r^{DNS}$',
    y_name=r'$\epsilon_r^{LES}$',
    )
f8.savefig(
    os.path.join(figures_folder, 'f8'), 
    transparent=True,
    bbox_inches='tight')
\end{lstlisting}

\noindent
The code above computes the residual dissipation rate obtained from the filtered DNS field and the corresponding prediction with the Smagorinsky model. The DNS-based value is evaluated directly from the definition of Eq.~\ref{eq:residual_dissipation_rate}, while the modelled contribution is obtained using the eddy-viscosity formulation with the prescribed Smagorinsky constant. The absolute error between the two fields is then computed and stored as an additional variable. Figure~\ref{fig:smagorinsky_model}.b displays the mid-plane distributions of $\epsilon_r^{DNS}$, the Smagorinsky prediction $\epsilon_r^{LES}$, and their absolute difference, together with the associated parity plot. Finally, the same procedure is applied to model the magnitude of the deviatoric residual stress. 

\begin{lstlisting}[style=mystyle, language=Python]
# Residual stress computation, reference and modeled
filtered_field.compute_tau_r(mode='DNS')
filtered_field.compute_tau_r(mode='Smag')

# Compute absolute error, save it to file, and add it as a variable
absolute_error = np.abs(filtered_field.TAU_r_DNS.value - filtered_field.TAU_r_Smag.value)
error_file_name = 'TAU_r_AE_id000.dat'
error_file_path = os.path.join(filtered_field.data_path, error_file_name)
ap.save_file(absolute_error, error_file_path)
ap.add_variable('TAU_r_AE', 'TAU_r_AE_{}.dat')
filtered_field.update(verbose=True)


# Residual sub-filter stress plot
f9, ax9 = filtered_field.plot_z_midplane(
    'TAU_r_DNS', 
    vmin=0, 
    vmax=10,
    title=r'$\tau_r^{DNS}$',
    transparent=False,
    scale='m',
    remove_cbar=True,
    colormap='Blues',
    )
f9.savefig(
    os.path.join(figures_folder, 'f9'), 
    transparent=True,
    bbox_inches='tight')

# Residual stress Smagorinsky plot
f10, ax10 = filtered_field.plot_z_midplane(
    'TAU_r_Smag', 
    vmin=-0, 
    vmax=10, 
    title=r'$\tau_r^{LES}$',
    transparent=False,
    scale='m',
    remove_y=True,
    remove_cbar=True,
    colormap='Blues'
    )
f10.savefig(
    os.path.join(figures_folder, 'f10'), 
    transparent=True,
    bbox_inches='tight')

# Absolute error plot
f11, ax11 = filtered_field.plot_z_midplane(
    'TAU_r_AE', 
    scale='m',
    colormap='Blues',
    title=r'$|\tau_r^{DNS} - \tau_r^{LES}|$',
    vmin=0,
    vmax=10,
    remove_y=True
    )
f11.savefig(
    os.path.join(figures_folder, 'f11'), 
    transparent=True,
    bbox_inches='tight')
# Parity plot
f12, ax12 = ap.parity_plot(
    filtered_field.TAU_r_DNS.value, 
    filtered_field.TAU_r_Smag.value, 
    density=True,
    cmin=1e-8,
    # limits=[-500, 500],
    # ticks=[-400, 400],
    rel_error=None,
    x_name=r'$\tau_r^{DNS}$',
    y_name=r'$\tau_r^{LES}$',
    limits=[0, 30],
    ticks=[0, 10, 20, 30]
    )
f12.savefig(
    os.path.join(figures_folder, 'f12'), 
    transparent=True,
    bbox_inches='tight')
\end{lstlisting}

The DNS reference and the Smagorinsky prediction are computed, and the absolute error between the two is stored as an additional variable. Figure \ref{fig:smagorinsky_model}.c shows the mid-plane distributions of $\tau_r^{DNS}$, the modeled value $\tau_r^{LES}$, and the corresponding absolute error, together with the associated parity plot. This final comparison demonstrates how the library enables \textit{a priori} assessment of sub-filter stress closures in addition to energy-transfer quantities, providing a comprehensive evaluation of turbulence models against DNS data.

\subsection{Machine-learning enhanced turbulent combustion closure}
In the present example we illustrate how to use the library to train a machine learning-based closure model for turbulent combustion. A multi-layer perceptron is trained on the filtered \textit{LES-like} data to regress the cell-reacting fraction parameter in the Partially-Stirred Reactor (PaSR) model. Specifically, the input parameters chosen are the Favre-filtered progress variable $\widetilde{c}$, its gradient module $\left|\nabla \widetilde{c}\right|$ and Laplacian $\nabla^2\widetilde{c}$, the chemical timescale $\tau_c$, and the mixing timescale $\tau_m$. The Favre-filtering operation is defined for a generic scalar $f$ as
\begin{equation}
    \widetilde{f} = \frac{\overline{\rho f}}{\overline{\rho}} \, .
\end{equation}
The selection of input parameters, baseline model, and the foundational idea of this tutorial is further detailed in previous research, where a similar approach was used to correct the PaSR model for methane premixed flames \cite{Piu2025}. The \texttt{NN} module of the library is used to build the training data tensor and generate a data scaler which can be fitted and used to extract the input data when testing the network on different cases. Finally, the results of the physics-based PaSR model and the deep-learning model are compared, and the effect of a sparsity regularisation term in the loss function is assessed both qualitatively and quantitatively.

The first part consists in importing the modules, creating the necessary folders, downloading the DNS data and initializing the field object. The simulation data are a subset of the hydrogen/air premixed turbulent flame described in \cite{Malé_Lapeyre_Noiray_2025} at equivalence ratio $\phi=0.5$, whose size was here reduced for reproducibility purposes. 

\begin{lstlisting}[style=mystyle, language=Python]
import numpy as np
import aPriori as ap
import os
import shutil

tmp_data_folder = 'tmp'
dns_data_folder = os.path.join(tmp_data_folder, 'DNS_data')

os.makedirs(tmp_data_folder, exist_ok=True)
os.makedirs(dns_data_folder, exist_ok=True)
if not os.path.exists(os.path.join(dns_data_folder, 'data')):
    ap.download(dataset='h2_premixed', dest_folder=dns_data_folder)

dns_field = ap.Field(dns_data_folder)
\end{lstlisting}

\begin{figure*}
    \centering
    \includegraphics[width=0.8\linewidth,trim=10 10 10 10,clip]{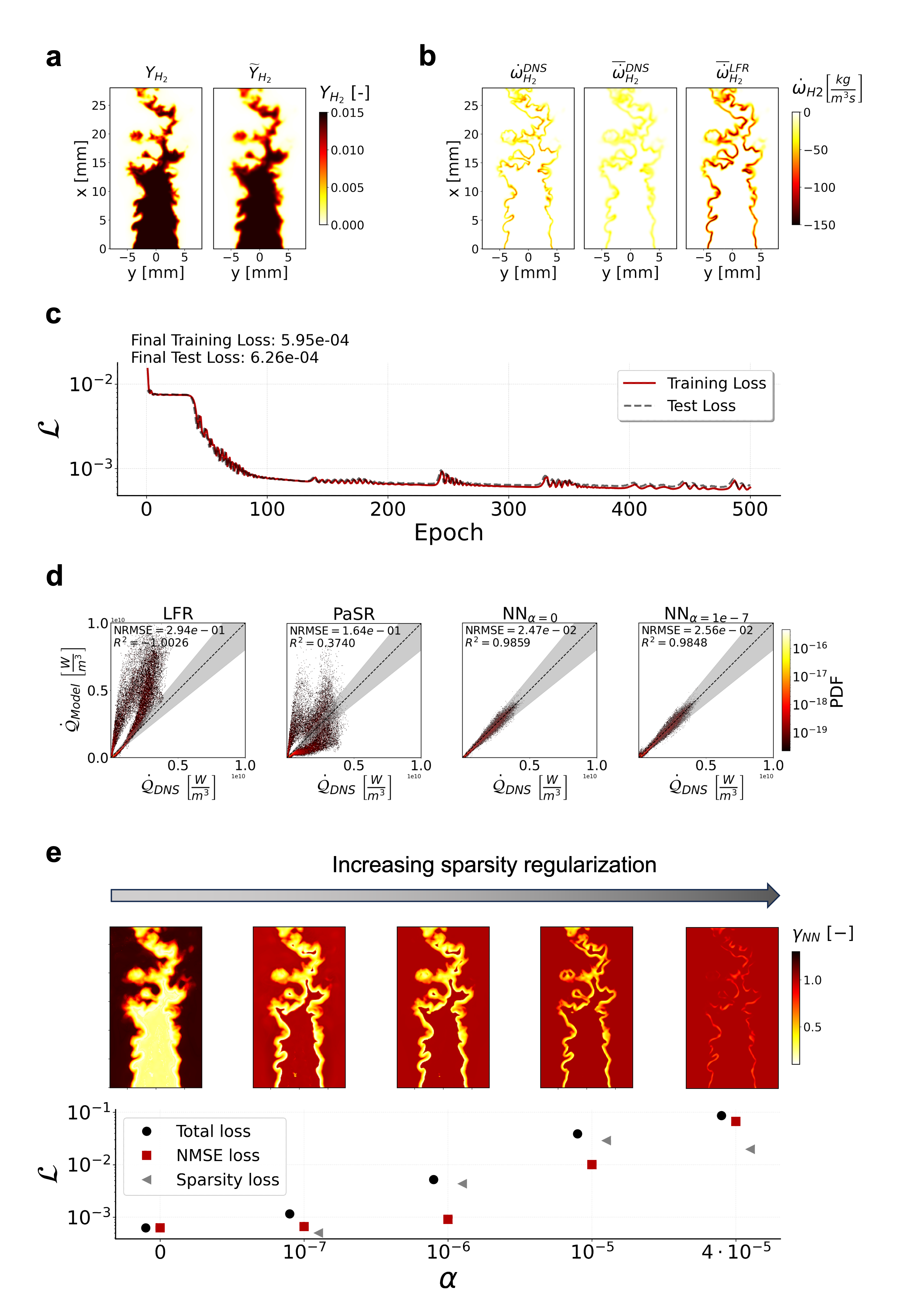}
    \caption{
    Machine-learning–assisted modelling of the cell-reacting fraction in a premixed hydrogen flame. 
    (a) DNS mass fraction of \(\mathrm{H_2}\) and its Favre-filtered counterpart, illustrating the effect of filtering on the thermochemical state. 
    (b) DNS reaction rate \(\dot{\omega}_{\mathrm{H_2}}^{\mathrm{DNS}}\), the corresponding filtered value \(\overline{\dot{\omega}}_{\mathrm{H_2}}^{\mathrm{DNS}}\), and the laminar finite-rate (LFR) estimate \(\dot{\omega}_{\mathrm{H_2}}^{\mathrm{LFR}}\). 
    (c) Training and test losses during optimisation of the neural-network model, showing rapid convergence and stable generalisation. 
    (d) Parity plots comparing DNS heat-release rate \(\dot{Q}_{\mathrm{DNS}}\) with (i) the LFR model, (ii) the PaSR closure, (iii) the neural network without regularisation (\(\alpha=0\)), and (iv) the neural network with weak sparsity regularisation (\(\alpha = 10^{-7}\)). 
    (e) Influence of increasing sparsity regularisation on the predicted reacting-fraction field \(\gamma_{\mathrm{NN}}\) (top) and on the contributions of total, NMSE, and sparsity losses (bottom). 
    }
    \label{fig:ml_closure}
\end{figure*}

\noindent
Subsequently, the reaction rates are computed on the DNS field, the field is then filtered, and the reaction rates are computed again on the filtered field. The reaction rates thus computed are referred to as Laminar Finite Rates, and the effect of filtering and computing the reaction rates is illustrated in Figure~\ref{fig:ml_closure}.a and Figure~\ref{fig:ml_closure}.b. The plotting function is available in the full tutorial and is omitted for simplicity.

\begin{lstlisting}[style=mystyle, language=Python]
# Compute reaction rates on the DNS
dns_field.compute_reaction_rates(exist_ok=True)

# Filter
filter_size = 6
filtered_field = ap.Field(dns_field.filter_favre(filter_size, exist_ok=True))

# Compute reaction rates on the filtered field
filtered_field.compute_reaction_rates(exist_ok=True)

plot_filtering_effect()
\end{lstlisting}

\noindent
The variables of interest are then computed on the \textit{LES-like} filtered field, with different methods that are identifiable by the root "\texttt{compute}". In order, we are computing the strain rate, the residual dissipation rate, the residual kinetic energy (necessary for the mixing timescale), the chemical and mixing timescale, the progress variable, its gradient and Laplacian.

\begin{lstlisting}[style=mystyle, language=Python]
# compute the strain rate on the filtered field
filtered_field.compute_strain_rate(save_tensor=True)

# compute the residual dissipation rate with Smagorinsky model
filtered_field.compute_residual_dissipation_rate(mode='Smag')

# compute residual kinetic energy
filtered_field.compute_residual_kinetic_energy()

# compute chemical timescale with SFR, FFR and Chomiak model
filtered_field.compute_chemical_timescale(mode='SFR', replace_nonreacting='max')
filtered_field.fuel = 'H2'
filtered_field.ox = 'O2'
filtered_field.compute_chemical_timescale(mode='Ch')
filtered_field.compute_mixing_timescale(mode='Kolmo')

# compute progress variable, its gradient, and the Laplacian
filtered_field.compute_progress_variable('H2O')
filtered_field.compute_gradient_C()
filtered_field.compute_laplacian_C()
\end{lstlisting}

\noindent
Once the derived variables are computed, the \texttt{Field} object can identify them as attributes, and they can be accessed by secondary classes. In the following script, we leverage the \texttt{TrainingBuilder} class, part of the \texttt{NN} module, which takes as input an object of the \texttt{Field} class and builds the tensor containing the input data for the neural network. This script extracts the mid-plane to speedup the training process for reproducibility purposes, then initializes a \texttt{TrainingBuilder} object to extract and scale the requested variables, and finally transfers the training data in a \texttt{pytorch}-compatible format. The script can also be run using the entire dataset (not only the mid-plane) by setting the flag \texttt{use\_midplane} to \texttt{False}.

\begin{lstlisting}[style=mystyle, language=Python]
use_midplane = True
train_size = 0.5
seed = 42
use_gpu = True

# Extract only the field midplane if requested
if use_midplane:
    filtered_field = ap.Field(filtered_field.cut([0, 0, 100],exist_ok=True))

# Process the data for the neural network
from aPriori.NN import TrainingBuilder
from architectures import select_device, prepare_training_data

# Initialize the scaler
scaler = TrainingBuilder()
scaler.add('C')
scaler.add('C_grad', mode='minmax')
scaler.add('C_laplacian', mode='minmax')
scaler.add('Tau_c_SFR', modulus=True, vmax=1e10, vmin=0, log=True)
scaler.add('Tau_m_Kolmo', modulus=True, log=True, vmin=1e-6)
scaler.fit(filtered_field)

# Build the training vector X and the reaction rates to use in the loss function
X = scaler.build_x(filtered_field)
HRR_LFR = filtered_field.HRR_LFR.reshape_column()
HRR_DNS = filtered_field.HRR_DNS.reshape_column()

# Save Scaler. This option is useful when testing on different datasets
scaler.save(os.path.join(tmp_data_folder, 'scaler.json'))

# Select the device to use (GPU)
device = select_device(cuda_device_id=0)
# Split the data between training and test, transfer in pytorch format
data = prepare_training_data(
    X, HRR_DNS, HRR_LFR, 
    train_size=train_size, 
    random_state=seed,
    device=device)
\end{lstlisting}

\noindent
The data processing done in the previous sections allows to use the data dictionary to train a neural network model using pytorch. The following section instantiates the network and uses a separately defined method for training. The loss function behaviour during the training epochs is plotted in Fig. \ref{fig:ml_closure}.c.

\begin{lstlisting}[style=mystyle, language=Python]
from architectures import FCNN, plot_training_loss
import torch

# Define hyperparameters
input_dim = X.shape[1]
n_layers = 5    # number of inner layers
tr_layers = 2   # transition layers
n_neurons = 64  # neurons in the inner layers
output_dim = 1
alpha = 0.0     # Sparsity-promoting regularization
epochs = 500
lr = 1e-3       # Learning rate
batch_size = 500000

model = FCNN(
    input_dim, 
    n_layers, 
    n_neurons, 
    output_dim, 
    tr_layers=tr_layers)

# Initialize model for reproducibility
model._initialize_weights(seed) 

model.fit(data, 
          epochs=epochs, 
          alpha=alpha,
          batch_size=batch_size
          )

plot_training_loss(
    model, 
    log_scale=True, 
    save_path=os.path.join(tmp_data_folder, 'loss.png'))
\end{lstlisting}

Finally, the model can be tested on the whole dataset. The model outputs are saved as files, and added as dynamic attributes with the \texttt{add\_attribute} function. This operation can be repeated considering different values of the regularization term $\alpha$. The results in Fig.~\ref{fig:ml_closure}.d illustrate with parity plots the neural network model calibration, comparing with the physics-based approach. Fig. \ref{fig:ml_closure}.e highlights the effect of the sparsity regularization, which for small values is able to reduce the activation of the deep-learning model limiting its effectiveness to the flame region, while for higher values of $\alpha$ the regularization term tends to overcome the NMSE loss, leading to an over-regularization of the model.

\begin{lstlisting}[style=mystyle, language=Python]
# Test on the whole dataset
model = model.to('cpu')
with torch.no_grad():
    gamma = model.forward(torch.tensor(X,dtype=torch.float32)).numpy()
HRR_FCNN = gamma*HRR_LFR

# Save output to read them as variables
gamma_path = 'gamma_fcnn_a{}_id000.dat'
gamma_attr = 'gamma_fcnn_a{}'
ap.save_file(gamma, os.path.join(filtered_field.data_path, gamma_path.format(0)))

# Update field to read
ap.add_variable(gamma_attr.format(0), gamma_path.format(0))
filtered_field.update()

torch.save(model, os.path.join(tmp_data_folder, f'model_a0.pth'))
\end{lstlisting}
          
\subsection{Computational Singular Perturbation (CSP) analysis}
The present section shows two main functionalities resulting from the coupling between the \texttt{aPriori} and the \texttt{PyCSP}~\cite{Malpica2022pycsp} packages. The main functionalities included in the current version of the library are the computation of the Tangential Stretching Rate (TSR), and the Amplitude Participation Indexes (APIs). The section is divided in three sub-sections, which briefly describe the theory behind the TSR, APIs, and a code example to compute these values with \texttt{aPriori}, respectively. 
The \texttt{PyCSP} software is described more in detail in \cite{Malpica2022pycsp}, and for a deeper discussion of the CSP, TSR, and APIs theory, the reader is referred to previous research work \cite{Lam1989csp, Valorani2015tsr, Valorani2020book}.

\subsubsection{Tangential stretching rate (TSR)} \label{subsubsec:tsr}

The tangential stretching rate provides a quantitative measure of the
rate at which neighbouring trajectories of the dynamical system stretch or
contract along the direction of the vector field.  
Given the homogeneous system
\begin{equation}
\frac{dz}{dt} = g(z),
\end{equation}
the evolution of an infinitesimal line element $v(t)$ joining two nearby
trajectories is governed by the linear variational equation
\begin{equation}
\frac{dv}{dt} = J_g(z)\,v,
\end{equation}
where $J_g$ is the Jacobian of the vector field.  
The instantaneous stretching rate along the direction of a unit vector
$\tilde{u}=v/\|v\|$ is given by the quadratic form
\begin{equation}
\omega_{\tilde{u}} = \tilde{u}\cdot J_g \cdot \tilde{u}.
\end{equation}

To characterize the stretching induced by the dynamics itself, the
direction $\tilde{u}$ is chosen to be tangent to the vector field,
\begin{equation}
\tilde{\tau}=\frac{g(z)}{\|g(z)\|},
\end{equation}
leading to the definition of the tangential stretching rate:
\begin{equation}
\omega_{\tilde{\tau}}=\tilde{\tau}\cdot J_g \cdot \tilde{\tau}.
\end{equation}

Expanding the vector field in the CSP eigenbasis,
\begin{equation}
g=\sum_{i=1}^N a_i f_i,
\end{equation}
and recalling that $J_g=A\,\Lambda\, B$ in this basis, the TSR admits the
modal representation
\begin{equation}
\omega_{\tilde{\tau}}=\sum_{i=1}^N W_i \lambda_i,
\end{equation}
where $\lambda_i$ is the eigenvalue associated with the $i$th mode and
\begin{equation}
W_i=\frac{f_i}{\|g\|}\left( \tilde{\tau} \cdot a_i \right)
\end{equation}
is a weight that quantifies the contribution of that mode.  
The TSR is thus a weighted average of the eigenvalues of $J_g$, combining
three ingredients:  
(i) the magnitude and sign of $\lambda_i$,  
(ii) the amplitude $f_i=b_i\cdot g$ of each mode,  
(iii) its degree of alignment with the vector field.

The TSR identifies the characteristic chemical time scale as
\begin{equation}
\tau_{\text{chem}}=\left|\frac{1}{\omega_{\tilde{\tau}}}\right|,
\end{equation}
which may be used to define Damköhler numbers or to detect explosive
($\omega_{\tilde{\tau}}>0$) versus dissipative ($\omega_{\tilde{\tau}}<0$)
regions of the flow.  
The TSR also induces a natural decomposition of the tangent space into fast,
active, and slow subspaces, reflecting which modes contribute most strongly
to the stretching dynamics \cite{Valorani2015tsr}.

\subsubsection{Amplitude Participation Indexes (APIs)}

The Amplitude Participation Index (API) quantifies the contribution of each
physical process (reaction or transport term) to the amplitude of a CSP
mode.  
For systems with transport and chemistry, the modal amplitudes are defined as
\begin{equation}
h_i = b_i \cdot (L(z) + g(z)),
\end{equation}
where $L(z)$ contains convective and diffusive operators and $g(z)$ the
chemical source term.  
Expanding $h_i$ into contributions from individual processes,
\begin{equation}
h_i = \sum_{l=1}^{N_{\text{proc}}} \beta_i^{(l)},
\end{equation}
where each $\beta_i^{(l)}$ corresponds to either a specific transport term or
a specific chemical reaction, the API is defined as
\begin{equation}
P_i^{(l)} =
\frac{\beta_i^{(l)}}{\sum_{j=1}^{N_{\text{proc}}} |\beta_i^{(j)}|}.
\end{equation}

By construction,  
\[
\sum_{l=1}^{N_{\text{proc}}} |P_i^{(l)}| = 1.
\]

The API identifies which processes dominate the behaviour of each mode.
For exhausted fast modes ($h_r\approx 0$), large values of $P_r^{(l)}$ reveal
the processes that enforce the local equilibrium condition expressed by
$h_r=0$.  
For slow active modes, the APIs provide insight into which reactions or
transport mechanisms govern the slow evolution of the state vector.

APIs form the basis for cause–effect analysis within CSP:  
they allow one to associate specific physical processes with specific modal
functions, enabling identification of the mechanisms responsible for slow
dynamics, equilibration, ignition, extinction, or transport–kinetics balance.
They also serve as building blocks for higher-level participation indices,
such as TSR-PIs or entropy-based indicators \cite{Valorani2020book}.

\subsubsection{CSP example}

The following example illustrates how the \texttt{aPriori} package can be used to
perform a local CSP analysis on a three–dimensional DNS dataset of a lifted \ce{H2}
jet flame. The script below loads the DNS field, computes CSP quantities including
the Tangential Stretching Rate (TSR) and the Amplitude Participation Indexes
(APIs), and produces a set of post–processing visualizations.

\begin{figure*}[ht!]
    \centering
    \includegraphics[width=1.0\linewidth]{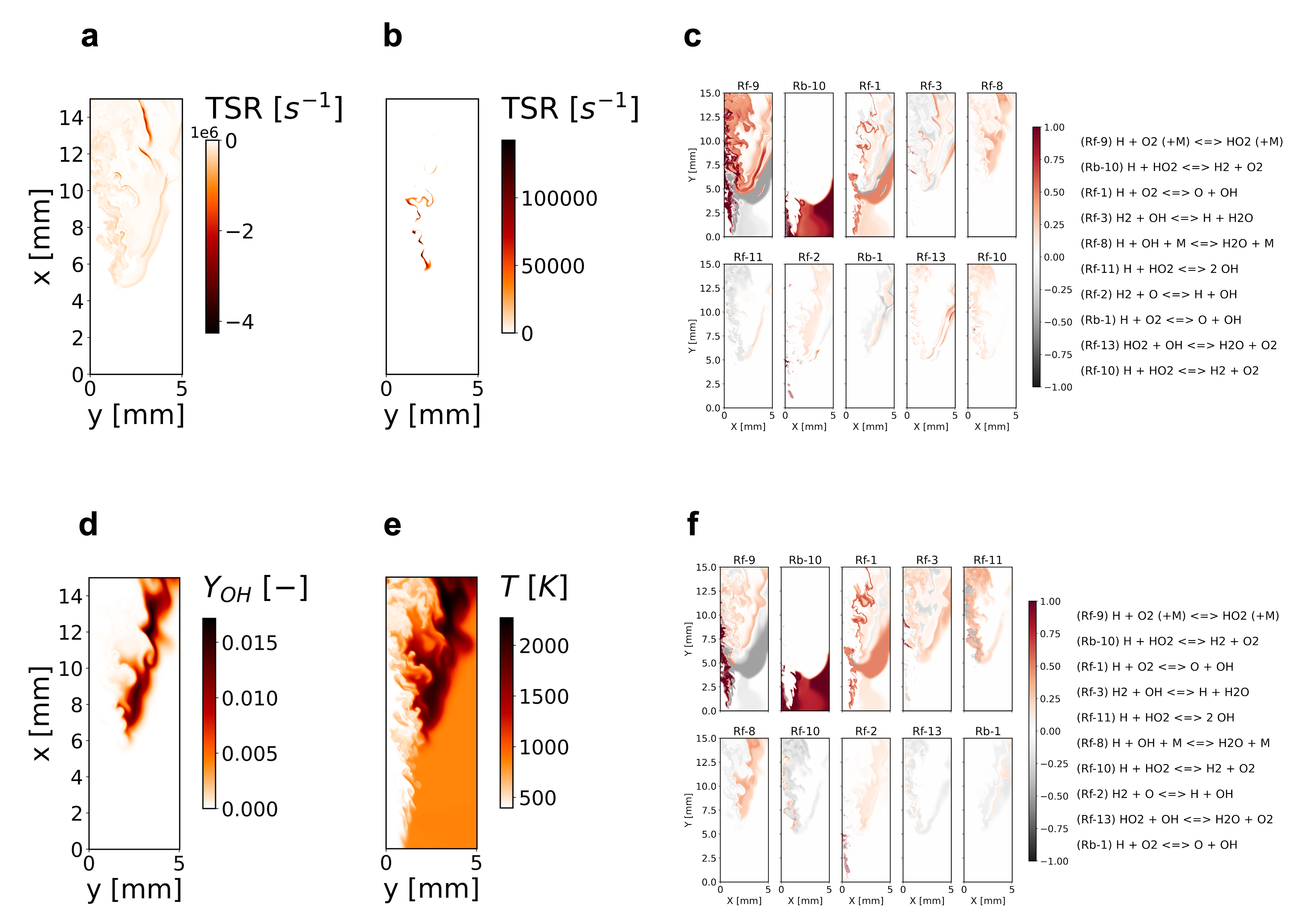}
    \caption{Illustration of CSP post-processing applied to the lifted \ce{H2} jet flame:
    (a) TSR field on the mid-plane, clipped on the negative values; (b) TSR field on the mid-plane, clipped on the positive values; (c) TSR amplitude participation indexes for ten
    dominant reactions; (d) OH mass fraction; (e) temperature; (f) OH-related amplitude participation
    indexes.}
    \label{fig:csp}
\end{figure*}

\begin{lstlisting}[style=mystyle, language=Python]
import aPriori as ap
import os

figures_folder = 'figures_csp'

# Load DNS dataset (PyCSP-compatible HDF5 structure)
dns_field = ap.Field('DNS_H2_lifted_csp_subdomain')

# Compute CSP quantities on the full 3D field
dns_field.compute_csp(
    API_species=['H2O', 'OH'], 
    n_chunks=100,
    )

os.makedirs(figures_folder, exist_ok=True)

# Plot modal participation in the TSR
dns_field.plot_api(
    api_type='TSR', 
    n=10, 
    n_cols=5, 
    cmap='RdGy_r', 
    vmin=-1, vmax=1,
    save_path=os.path.join(figures_folder, 'api_tsr.png')
    )

# Plot API for OH production/consumption
dns_field.plot_api(
    api_type='OH', 
    n=10, 
    n_cols=5, 
    cmap='RdGy_r', 
    vmin=-1, vmax=1,
    save_path=os.path.join(figures_folder, 'api_oh.png')
    )

# Mid-plane map of the TSR field
dns_field.plot_z_midplane(
    'TSR_DNS',
    colormap='gist_heat',
    transpose=True,
    remove_title=True,
    cbar_title='TSR $[s^{-1}]$',
    vmax=0,
    transparent=False,
    save_path=os.path.join(figures_folder, 'TSR.png')
    )

# Mid-plane map of the OH mass fraction
dns_field.plot_z_midplane(
    'YOH',
    colormap='gist_heat_r',
    transpose=True,
    remove_title=True,
    cbar_title=r'$Y_{OH}$ $[-]$',
    save_path=os.path.join(figures_folder, 'YOH.png')
    )
\end{lstlisting}

Figure~\ref{fig:csp} summarizes the outcome of the CSP analysis applied to the lifted \ce{H2} jet flame. Figure~\ref{fig:csp}.a reports the spatial distribution of the negative Tangential Stretching Rate (TSR) values, while Figure~\ref{fig:csp}.b highlights only its positive values, which identify regions where the local chemical dynamics is explosive. These regions are spatially confined and closely aligned with the flame front, indicating that positive TSR values only arise where chain-branching chemistry overcomes dissipative relaxation.

The APIs contributing to the TSR are shown in Figure~\ref{fig:csp}.c for the ten most relevant reactions, which relative importance is selected based on the $\ell^1$ norm of its associated API computed on the whole domain. The dominant contribution in regions of high positive TSR is associated with the chain-branching step
\vspace{2pt}
\\
$\text{(Rf-1)}\quad \ce{H + O2 <=> O + OH},$
\vspace{2pt}
\\
which is known to control the explosive growth of radical pools in hydrogen oxidation. This reaction exhibits strong positive participation precisely in the regions where TSR is positive, confirming its leading role in setting the most energetic chemical timescale.

A second set of reactions involves pathways associated with the \ce{HO2} chemistry, but their roles are markedly different. The reaction
\vspace{2pt}
\\
$\text{(Rf-9)}\quad \ce{H + O2 (\text{+}M) <=> HO2 (\text{+}M)}$
\vspace{2pt}
\\
exhibits significant participation in regions adjacent to the flame front, where \ce{H} radicals and molecular oxygen coexist. Its activity extends into the preheat layer, reflecting its role in regulating radical pool formation under conditions where chain branching is not yet dominant. In contrast, the reverse reaction
\vspace{2pt}
\\
$\text{(Rb-10)}\quad \ce{H + HO2 <=> H2 + O2}$
\vspace{2pt}
\\
is found to contribute primarily in regions characterized by weak or absent heat release, upstream of the main reaction zone. There, it acts as a radical recombination pathway in partially mixed or chemically frozen regions, where \ce{H2} and \ce{O2} are not well mixed yet and radicals are almost absent. Accordingly, while \mbox{(Rb-10)} displays a large amplitude participation index, its contribution is largely decoupled from the explosive dynamics identified by positive TSR values.

We can identify a third group of chain-propagation and termination reactions, such as
\vspace{2pt}
\\
$\text{(Rf-3)}\quad \ce{H2 + OH <=> H + H2O}$, and
\vspace{2pt}
\\
$\text{(Rf-8)}\quad \ce{H + OH + M <=> H2O + M}$,
\vspace{2pt}
\\
which show weaker and more diffuse participation. Their role is primarily dissipative, contributing to the relaxation toward the slow manifold once the radical pool has been established. Notably, \mbox{(Rf-8)} exhibits a clear correlation with the high temperature region (Figure~\ref{fig:csp}.e), consistent with its role as a three-body recombination process that becomes active during thermal relaxation and radical quenching.

Figure~\ref{fig:csp}.d reports the OH mass fraction, while Figure~\ref{fig:csp}.e shows the temperature field. Both quantities peak in the same region where the TSR exhibits its largest magnitude, confirming the strong coupling between radical production, heat release, and the emergence of fast chemical time scales.

Finally, Figure~\ref{fig:csp}.f shows the APIs associated with the OH source term. In this case, the dominant contributions arise from reactions directly producing or consuming OH, most notably the chain-branching step \mbox{(Rf-1)} and the propagation reaction \mbox{(Rf-3)} assume slightly higher APIs in the preheating zone compared to the TSR-based APIs. Notably, also the relative importance of \mbox{(Rf-11)} increases. Compared to the TSR-based APIs, the OH APIs are more spatially localized and closely follow the thin reaction layer, highlighting the distinction between reactions governing a determined species evolution and those controlling in general the fast modes of the system.

Overall, this example demonstrates how CSP-based indicators provided by \texttt{aPriori} allow one to (i) identify explosive versus dissipative regions via the TSR, (ii) isolate the chemical pathways responsible for the most energetic time scales, and (iii) distinguish between reactions controlling a selected species production and those governing thermochemical relaxation. While this example used a relatively simple and well known system, this functionality can be leveraged in various scenarios, including detailed analysis of pollutants or greenhouse gases formation pathways in complex combustion simulations. These results stem from the interaction between two community-driven, open-source software projects, illustrating how coordinated and openly shared development efforts can facilitate the practical adoption of advanced analysis techniques while reducing implementation overhead.

\section{Conclusions and future development}\label{sec:conclusions}

The \texttt{aPriori} package addresses a critical gap in the post-processing of direct numerical simulation data by offering a lightweight, memory-efficient, and modular Python-based solution. By leveraging a pointer-based data access strategy, the software reduces memory consumption at the cost of reading the file each time its content is accessed; the reading time is shown to be in the order of seconds for hundred millions of grid points. This trade-off enables the processing of large-scale simulations without the need for high-performance computing infrastructure, thus significantly broadening access to DNS data.

This initiative bridges a longstanding gap between traditional CFD solvers—typically written in low-level languages such as C++ or Fortran—and modern data-driven workflows that rely heavily on Python-based machine learning and statistics libraries. By doing so, \texttt{aPriori} facilitates a smoother transition from raw simulation data to advanced modeling and analysis pipelines, enabling new synergies between high-fidelity physics and data-driven approaches. Additionally, the implementation of data chunking enables memory-safe evaluations of complex thermochemical expressions, a crucial feature for combustion applications involving detailed chemistry, such as source term reconstruction and Computational Singular Perturbation (CSP) analyses.

\texttt{aPriori} also contributes to the broader research ecosystem. By providing a standardized, open-source toolkit to process complex and heterogeneous DNS datasets, the package encourages data reuse and reproducibility, facilitating data sharing across institutions. In particular, its compatibility with the BLASTNet platform enhances the usability of publicly available datasets and reduces the need for redundant, resource-intensive simulations.
Additionally, the Python interface ensures seamless integration with a wide variety of scientific libraries—including Cantera, Matplotlib, and PyVista—promoting interoperability and ease of adoption.

Several development paths can further enhance the capabilities of the library. These include the implementation of more diverse and flexible gradient computation schemes, the addition of 3D visualization utilities, and support for more abstract data types such as vectors and tensors to enhance the software's modularity. Finally, the authors would like to stress that, as an open-source project, \texttt{aPriori} is designed to evolve with the needs of the community. Its design invites contributions and encourages a user-driven roadmap, aiming, with cooperation within the community, to become a reference tool for the post-processing of direct numerical simulations.




\bibliographystyle{elsarticle-num} 
\bibliography{references}



\end{document}